\documentclass[10pt, journal, twoside]{IEEEtran}
\normalsize

\usepackage{amsmath}
\usepackage{amssymb}
\usepackage{amsfonts}
\usepackage{graphicx}
\usepackage{epstopdf}
\usepackage{subfigure} 
\usepackage{color}
\usepackage{bm}
\usepackage{url}
\usepackage[ruled,boxed,linesnumbered,commentsnumbered]{algorithm2e}
\usepackage{setspace}
\usepackage{cite}
\usepackage{upgreek}
\usepackage[colorlinks,linkcolor=blue,anchorcolor=blue,citecolor=blue]{hyperref}

\ifCLASSINFOpdf
\else
\fi

\begin{document}
\title{Joint Channel Estimation and Prediction for Massive MIMO with Frequency Hopping Sounding}

\author{\normalsize{Yiming~Zhu,~\IEEEmembership{Student Member,~IEEE},
Jiawei~Zhuang, 
Gangle~Sun,~\IEEEmembership{Student Member,~IEEE}, 
Hongwei~Hou,~\IEEEmembership{Graduate Student Member,~IEEE}, 
Li~You,~\IEEEmembership{Senior Member,~IEEE}, 
and 
Wenjin~Wang,~\IEEEmembership{Member,~IEEE}} % <-this % stops a space
\thanks{Manuscript received xxx; revised xxx. \textit{(Yiming Zhu and Jiawei Zhuang contributed equally to this work.) (Corresponding author: Wenjin Wang.)}}% <-this % stops a space
\thanks{Yiming~Zhu, Jiawei~Zhuang, Gangle~Sun, Hongwei~Hou, Li~You, and Wenjin~Wang are with the National Mobile Communications Research Laboratory, Southeast University, Nanjing 210096, China, and also with Purple Mountain Laboratories, Nanjing 211100, China (e-mail: ymzhu@seu.edu.cn; jw-zhuang@seu.edu.cn; sungangle@seu.edu.cn; hongweihou@seu.edu.cn; lyou@seu.edu.cn; wangwj@seu.edu.cn).}
}


\markboth{IEEE Transactions on Communications}%
{Submitted paper}



\maketitle

\begin{abstract}
%\boldmath  

In massive multiple-input multiple-output (MIMO) systems, the downlink transmission performance heavily relies on accurate channel state information (CSI). Constrained by the transmitted power, user equipment always transmits sounding reference signals (SRSs) to the base station through frequency hopping, which will be leveraged to estimate uplink CSI and subsequently predict downlink CSI.
This paper aims to investigate joint channel estimation and prediction (JCEP) for massive MIMO with frequency hopping sounding (FHS). 
Specifically, we present a multiple-subband (MS) delay-angle-Doppler (DAD) domain channel model with off-grid basis to tackle the energy leakage problem. Furthermore, we formulate the JCEP problem with FHS as a multiple measurement vector (MMV) problem, facilitating the sharing of common CSI across different subbands. 
To solve this problem, we propose an efficient Off-Grid-MS hybrid message passing (HMP) algorithm under the constrained Bethe free energy (BFE) framework. 
Aiming to address the lack of prior CSI in practical scenarios, the proposed algorithm can adaptively learn the hyper-parameters of the channel by minimizing the corresponding terms in the BFE expression. 
To alleviate the complexity of channel hyper-parameter learning, we leverage the approximations of the off-grid matrices to simplify the off-grid hyper-parameter estimation. 
Numerical results illustrate that the proposed algorithm can effectively mitigate the energy leakage issue and exploit the common CSI across different subbands, acquiring more accurate CSI compared to state-of-the-art counterparts.

\end{abstract}


% \fi
\IEEEpeerreviewmaketitle

\section{Introduction}

% massive MIMO-OFDM
\IEEEPARstart{M}{A}{S}{S}{I}{V}{E} multiple-input multiple-output orthogonal frequency division multiplexing (MIMO-OFDM) is widely used in current communication systems due to its notable advantages, encompassing high spectral and energy efficiency, high data rate, and strong robustness to frequency-selective fading \cite{bjornson2016massive,li2016statistical,you2016channel,you2017bdma,zaidi2016waveform}.
In massive MIMO-OFDM systems, spatial multiplexing techniques, such as precoding \cite{alodeh2018symbol} and beamforming \cite{molisch2017hybrid}, heavily rely on accurate channel state information (CSI). 
However, the limited pilot overhead in practical systems poses significant challenges to accurate CSI acquisition.

In time-division duplexing (TDD) massive MIMO-OFDM systems, to enhance the accuracy of CSI acquisition with the limited pilot overhead, CSI acquisition can be divided into two steps: estimating the uplink pilot channels based on sounding reference signals (SRSs) \cite{38.211}, followed by predicting the downlink data channels based on the estimated uplink pilot channels to resolve the transmission issues caused by CSI staleness. 
Limited by user equipment (UE) transmitted power, the base station (BS) prefers to configure the UE to sound the channel with frequency hopping \cite{zhao2017system,dahlman20205g,jin2023massive,38.211}, significantly amplifying the received power compared to the fullband SRS transmission. 
However, the FHS mode restricts the BS to estimating only a subset of the fullband channel in a single SRS transmission, thereby putting forward higher requirements for fullband CSI acquisition. 
This paper considers the frequency hopping SRS transmission with limited pilot overhead and aims to design efficient joint channel estimation and prediction (JCEP) algorithms.

\subsection{Prior Work and Motivations}
Due to the limited local scatterers, the realistic channels exhibit inherent sparsity in massive MIMO-OFDM systems\cite{wang2018survey}. 
For this reason, various channel estimation schemes exploiting channel sparsity have been extensively investigated in the literature for massive MIMO-OFDM systems. 
Building on the delay-angle domain sparsity of massive MIMO-OFDM channels, 
the authors in \cite{you2016channel} developed a channel estimation approach based on adjustable phase shift pilots. 
Based on the proposed concept of the joint delay-angle subspace, a sparse signal recovery method was developed for efficient sparse channel estimation \cite{2016WangDongMing}. 
By exploiting the sparse nature of millimeter-wave (mmWave) channels, an efficient open-loop channel estimator based on the orthogonal matching pursuit (OMP) algorithm was developed for an mmWave hybrid MIMO system \cite{lee2016channel}. 
Statistical inference methods have recently been developed for efficient sparse channel recovery. 
The expectation-maximization (EM) forms of the expectation propagation (EP) \cite{MessagePassingReceiver2016}, generalized approximate message passing (AMP), and vector AMP algorithms \cite{BroadbandMillimeterWave2018} exploit the channel sparsity to estimate the massive MIMO-OFDM channels.
Based on the constrained Bethe free energy (BFE) framework, a sparse channel estimation scheme in \cite{liu2021sparse} was proposed, which captures the sparse structure and temporal dependency of massive MIMO-OFDM channels using a hidden Markov probability model. 
In \cite{wan2023robust,wan2024two}, CSI acquisition schemes considering imperfect factors are proposed by exploiting the dynamic sparsity of massive MIMO-OFDM channels, paving the way for accurate CSI acquisition in practical TDD communication systems.

% why CP
In practical communication systems, if the estimated uplink pilot channels are directly used in the spatial multiplexing techniques for the downlink data transmission, it may face the challenge of CSI staleness provided the short coherence time against the CSI delay, especially in high-mobility scenarios. To address the issue of CSI staleness, most of the relevant work paid great attention to the channel prediction technique, which predicts the future downlink data channels by exploiting the temporal correlation of the past estimated pilot channels.
A variety of channel prediction techniques for single-input single-output systems have been studied in the literature, e.g., the autoregressive (AR) model-based methods \cite{duel2007fading} and the sum-of-sinusoids (SOS) model-based methods \cite{wong2006wlc43}.
Recently, the channel prediction technique has been extended to the massive MIMO-OFDM systems.
Equipped with large-scale antenna arrays, massive MIMO-OFDM systems create new opportunities for channel prediction. 
Due to the stationary scattering environment, there exists a high correlation between adjacent subcarriers and antennas, which can be further decoupled to high-resolution delay-angle domain channels via inverse discrete Fourier transform (IDFT) \cite{li2016statistical,you2016channel,you2017bdma}. 
Different from the conventional methods, which only leverage temporal correlation, various channel prediction methods have been proposed for massive MIMO-OFDM systems using frequency, spatial, and temporal correlation \cite{lv2019channel,yin2020addressing,qin2022partial,wu2021channel}. 
The authors in \cite{lv2019channel} analyzed the impacts of channel representations in different domains on channel prediction and proposed a low-complexity AR-based channel predictor using the sparsity of delay-angle domain channels in mmWave MIMO-OFDM systems. 
To deal with the ``curse of mobility'', a Prony-based delay-angle domain (PDA) channel prediction method was proposed, which utilizes the specific delay-angle-Doppler (DAD) structure of the multipath \cite{yin2020addressing}. 
In \cite{qin2022partial}, a joint DAD wideband channel prediction method was proposed, which extracts the Doppler shifts by matrix pencil (MP) method for channel prediction. 
By leveraging the residual temporal correlation between the neighboring channel elements introduced by the energy leakage, a spatial-temporal AR (ST-AR)-based channel prediction method was proposed in \cite{wu2021channel}.

% Motivation 

From the state-of-the-art overview, the aforementioned works treat channel estimation and prediction as separate modules. 
In essence, channel estimation and prediction can be classified as CSI acquisition, which mainly differ in the time-domain location of CSI. 
Specifically, channel estimation aims to acquire the current CSI on the uplink pilot symbols, while channel prediction intends to acquire the future CSI on the downlink data symbols. 
Therefore, channel estimation and prediction can be jointly operated to enhance CSI acquisition accuracy by utilizing the sparsity of massive MIMO-OFDM channels.
Moreover, previous research has primarily concentrated on CSI acquisition with fullband sounding, which may suffer performance deterioration in the systems with FHS, thus spurring our research on JCEP in such scenarios.

\subsection{Main Contributions}
Motivated by the considerations outlined above, this paper aims to design efficient JCEP algorithms utilizing the characteristics of FHS. 
Our contributions are summarized as follows:

\begin{itemize}
	\item We prove that the resolution of the sparsity-domain channel based on DFT is proportional to the number of channel samples. Considering the limited number of channel samples in the frequency-space-time (FST) domain in the practical scenarios, we analyze that the energy leakage problem of the DAD domain channel based on DFT is inevitable. For this reason, we propose a multiple-subband (MS) DAD-domain channel model with off-grid basis to address the energy leakage problem.
	\item We reveal that the channels of different subbands are independently and identically distributed (i.i.d.), allowing us to formulate the JCEP problem with FHS as a generalized multiple measurement vector (MMV) problem. 
	Thus, the common channel statistical characteristics of different subbands can be exploited to enhance the accuracy of CSI acquisition. 
	To achieve efficient JCEP, we approximate the MMV problem as a BFE minimization problem with hybrid constraints to balance tractability and fidelity. 
	\item Based on the BFE minimization with hybrid constraints, we propose an efficient Off-Grid-MS hybrid message passing (HMP) algorithm to solve the JCEP problem with FHS. To deal with the lack of prior CSI in practical
	scenarios, we utilize the proposed algorithm to adaptively learn the hyper-parameters, which vary with the channel scenario and are typically unknown to the BS. Additionally, we further reduce the computational complexity of off-grid hyper-parameters by leveraging the approximations of the off-grid matrices.
\end{itemize}

The remainder of this paper is organized as follows: 
Section~\ref{sec2} introduces the system model. 
Section \ref{sec3} presents the JCEP problem formulation. 
Section \ref{sec4} starts from the constrained BFE minimization framework and develops the JCEP algorithm with FHS. 
Simulation results are provided in Section \ref{sec5}, and conclusions are drawn in Section \ref{sec6}.

\emph{Notations:} We use $ \bar{\jmath}=\sqrt{-1} $ to denote the imaginary unit. 
The expression $ \mathcal{CN}(x;\mu,\tau) $ denotes the circularly symmetric complex Gaussian distribution of variable $ x $ with mean $ \mu $ and variance $ \tau $. 
$[\cdot]_{i}$ and $[\cdot]_{i,j}$ denote the $i$-th element of a vector and the $(i,j)$-th element of a matrix, respectively.
$ \text{E}[\cdot] $ and $ \text{V}[\cdot] $ are the expectation and variance operations, respectively. 
$ \text{Tr}\lbrace\cdot\rbrace $ denotes the trace operation.
$ \text{Re}\lbrace \cdot \rbrace $ takes the real part of complex variable. 
$ \mathbf{I}_{M} $ represents the $ M\times M $ dimensional identity matrix.  
$ \mathbf{1}_{N} $ and $ \mathbf{0}_{N} $ denote the $ N $ dimensional all-one and all-zero vectors, respectively.  
$ \text{diag}\lbrace\mathbf{x}\rbrace $ represents a diagonal matrix with main diagonal $ \mathbf{x} $. 
Scalars, column vectors, and matrices are represented by lowercase, bold lowercase, and bold uppercase, respectively. 
The superscripts $ (\cdot)^{T} $, $ (\cdot)^{*} $, and $ (\cdot)^{H} $  denote the transpose, conjugate, and conjugate-transpose of a vector or matrix, respectively. 
The real number, complex number, integer, and binary fields are represented by the symbols $ \mathbb{R} $, $ \mathbb{C} $, $ \mathbb{Z} $, and $\mathbb{B}$, respectively. 
The symbol $ \otimes $ denotes the Kronecker product. 
The Dirac and Kronecker delta functions are represented by $ \delta(\cdot) $ and $ \delta[\cdot] $, respectively. 
Further, $ \Vert\cdot\Vert_{\text{F}} $ is the Frobenius norm.

\section{System Model}
\label{sec2}
This paper considers a TDD massive MIMO-OFDM system. 
The BS with a uniform planar array (UPA)  serves multiple UEs with an omnidirectional antenna. The UPA with $ M=M_{\text{v}}M_{\text{h}} $ antennas comprises $ M_{\text{v}} $ and $ M_{\text{h}} $ antennas in vertical and horizontal directions, respectively, with one-half wavelength spacing. 
Before transmission, the signal is modulated by OFDM comprising $N_{\text{FFT}}$ subcarriers spaced by $ \Delta \phi $ and $N_{\text{CP}}$-length cyclic prefix (CP). The number of subcarriers for data transmission is $ N_{\text{SC}} $.
Hence, the system sampling interval, the OFDM symbol duration, and the CP duration are $ T_{\text{sam}} = \frac{1}{N_{\text{FFT}}\Delta \phi} $, $ T_{\text{sym}}=N_{\text{FFT}}T_{\text{sam}} $, and $ T_{\text{CP}}=N_{\text{CP}}T_{\text{sam}} $, respectively. 

\begin{figure}[!t]
	\centering
	\includegraphics[width=\linewidth]{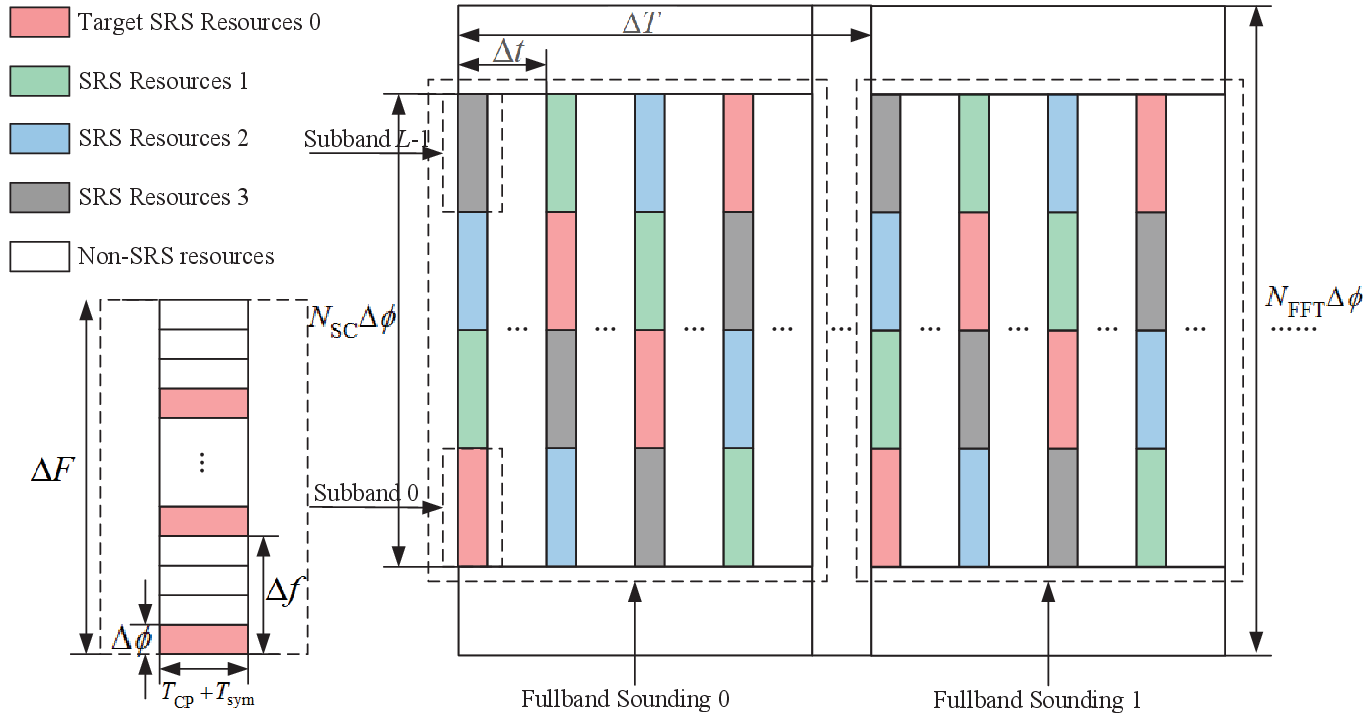}
	\caption{SRS configuration.}
	\label{fig:SRS configure}
\end{figure}

Thanks to the channel reciprocity of the TDD systems, the BS can measure the downlink CSI based on the uplink SRS. If the fullband sounding is completed by only one SRS transmission, the BS received power will be low due to the limited UE transmitted power, resulting in poor CSI accuracy. To improve the BS received power, we consider the FHS, a transmission mode specified by the 3rd Generation Partnership Project (3GPP) standards \cite{38.211}, as depicted in Fig.~\ref{fig:SRS configure}. 
In FHS mode, the available bandwidth is divided into $L$ subbands, each with a subband spacing of $ \Delta F = \frac{N_{\text{SC}}\Delta \phi}{L} $. 
The transmission comb number of pilot symbols is $ N_{\text{TC}} $ for frequency multiplexing, so each SRS transmission sounds $ N=\frac{N_{\text{SC}}}{LN_{\text{TC}}} $ subcarriers of one subband in one OFDM symbol, with a subcarrier spacing of $ \Delta f = N_{\text{TC}}\Delta \phi $. 
It is assumed that each UE sounds the fullband channel $ K $ times through $ KL $ SRS transmissions, and the $ (kL+l) $-th SRS transmission corresponds to the $ q_{l} $-th subband $ (0\le q_{l}<L) $ channel.
In the time domain, the time interval between two adjacent SRS transmissions and the time interval between two adjacent fullband soundings are denoted as $ \Delta t $ and $ \Delta T $, respectively. 

\subsection{Signal Model}
Considering that the target SRS resources are orthogonal to other SRS resources by frequency, time, and code division multiplexing, we can pay attention to the signal processing on the target SRS resources of one single UE and process the signals on other SRS resources synchronously in the same manner \cite{dahlman20205g}. Consequently, we can express the received signal on the $l$-th subband $ \mathbf{y}_{l}\in\mathbb{C}^{NMK} $ at the BS as 
\begin{align}
	\label{eq:received signal model}
	\mathbf{y}_{l} &= \mathbf{S}\mathbf{g}_{l} + \mathbf{z}_{l},
\end{align}
whose the $(nMK+mK+k)$-th element $ y_{nmkl} $ represents the received signal between the UE and the $ m $-th BS antenna on the $ n $-th subcarrier of the $ (kL+l) $-th SRS resource. 
$ \mathbf{S} \triangleq \text{diag}\lbrace \mathbf{s} \rbrace $ is the pilot matrix, and the $ (nMK + mK + k) $-th element of $ \mathbf{s}\in\mathbb{C}^{NMK} $ is $ s_{nmk} $. 
Note that the power of pilot symbols equals $ 1 $, i.e., $ \mathbf{S}\mathbf{S}^{H} = \mathbf{I}_{NMK} $. 
The vector $\mathbf{g}_{l}$ denotes the frequency-space-time-domain channel response vector (FSTCRV), whose specific form will be discussed in Section \ref{sec:channel model}. 
The vector $ \mathbf{z}_{l}$ represents the complex Gaussian noise vector on the $l$-th subband, with its $ (nMK + mK + k) $-th element following the i.i.d. complex Gaussian distribution $ \mathcal{CN}(z_{nmkl};0,\sigma_{\text{z}}) $. 

\subsection{Channel Model}\label{sec:channel model}
We consider a massive MIMO-OFDM channel where transmitted signals reach received antennas through multipath propagation. To specify one particular path, we define the DAD tuple as $ (\tau,\theta,\varphi,\nu) $, where $ \tau\in(0,\tau_{\text{max}}] $, $ \theta\in[0,\pi] $, $ \varphi\in[0,\pi] $, $ \nu\in(\nu_{\text{min}},\nu_{\text{max}}] $ denote the delay, elevation angle of arrival (AoA), azimuth AoA, and Doppler shift, respectively. 
Following the channel modeling in \cite{wu2021channel}, we construct one virtual path by merging the physical paths with similar DAD tuples, thereby grouping and partially decoupling physical paths.

After the construction of virtual paths, we define the complex gain of one specific path between the UE and the BS as the function of the DAD tuple, denoted as $ a(\tau,\theta,\varphi,\nu) $. We assume the multipath channel follows uncorrelated Rayleigh fading, where different paths are independent. Thus, the complex gain $ a(\tau,\theta,\varphi,\nu) $ satisfies \cite{you2016channel,ma2018sparse}
\begin{align}
	\label{eq:E complex gain}
	\text{E}&\left[  a(\tau,\theta,\varphi,\nu)a^{*}(\tau^{\prime},\theta^{\prime},\varphi^{\prime},\nu^{\prime}) \right] \nonumber \\
	&=A(\tau,\theta,\varphi,\nu)\delta(\tau-\tau^{\prime})\delta(\theta-\theta^{\prime})
	\delta(\varphi-\varphi^{\prime})\delta(\nu-\nu^{\prime}),
\end{align}
where $ A(\tau,\theta,\varphi,\nu) $ represents the power DAD spectrum between the UE and the BS.

Compared with the fading coefficients, physical channel parameters, such as delays, angles, and Doppler shifts, vary much more slowly. 
Before presenting the channel model, we introduce the assumption that the stationary time, i.e., the period with constant physical parameters, is longer than the sounding time $ K\Delta T $, as mentioned in \cite{qin2022partial}. 
Therefore, the scattering environment remains almost unchanged, allowing physical channel parameters to be treated as constants relative to fading coefficients during the sounding time.

On the above basis, the FSTCRV on the $ l $-th subband $ \mathbf{g}_{l} $ can be denoted as \cite{you2016channel,ma2018sparse}
\begin{align}
	\label{eq:FSTCRV}
	\mathbf{g}_{l}\hspace{-1mm}=\hspace{-2mm}\int_{0}^{\tau_{\text{max}}}\hspace{-2mm}\int_{0}^{\pi}\hspace{-2mm}\int_{0}^{\pi}\hspace{-2mm}\int_{\nu_{\text{min}}}^{\nu_{\text{max}}}\hspace{-2mm}a(\tau,\theta,\varphi,\nu)
	\mathbf{w}(\tau,\theta,\varphi,\nu)\psi_{l}(\tau,\nu)
	\text{d}\nu\text{d}\varphi\text{d}\theta\text{d}\tau,
\end{align}
whose the $ (nMK+mK+k) $-th element $g_{nmkl}$ is the FSTCR between the UE and the $ m $-th BS antenna on the $ n $-th subcarrier of the $ (kL+l) $-th SRS resource. 
Note that the $ m $-th BS antenna corresponds to the antenna with vertical index $m_{\text{v}}$ and horizontal index $m_{\text{h}}$, denoted as $ m\triangleq m_{\text{v}}M_{\text{h}}+m_{\text{h}} $. 
For brevity, we define $ \mathbf{w}(\tau,\theta,\varphi,\nu)\triangleq \mathbf{b}(\tau)\otimes \mathbf{c}_{\text{v}}(\theta)\otimes 
\mathbf{c}_{\text{h}}(\varphi;\theta)\otimes \mathbf{d}(\nu)  $. 
The vectors $ \mathbf{b}(\tau)\in\mathbb{C}^{N} $, $ \mathbf{c}_{\text{v}}(\theta)\in\mathbb{C}^{M_{\text{v}}} $, $ \mathbf{c}_{\text{h}}(\varphi;\theta)\in\mathbb{C}^{M_{\text{h}}} $, and $ \mathbf{d}(\nu)\in\mathbb{C}^{K} $ represent the steering vectors in the delay, elevation angle, azimuth angle, and Doppler domains \cite{yin2020addressing,wu2021channel}. They are defined as $[\mathbf{b}(\tau)]_{n} \triangleq e^{-\bar{\jmath}2\pi n\Delta f\tau}$, $[\mathbf{c}_{\text{v}}(\theta)]_{n} \triangleq e^{-\bar{\jmath}\pi n\cos\theta} $, $[\mathbf{c}_{\text{h}}(\varphi;\theta)]_{n} \triangleq e^{-\bar{\jmath}\pi n\sin\theta\cos\varphi} $, and $ [\mathbf{d}(\nu)]_{n} \triangleq e^{\bar{\jmath}2\pi n\Delta T\nu} $, respectively. 
The phase difference of the $l$-th subband is defined as $\psi_{l}(\tau,\nu) \triangleq e^{\bar{\jmath}2\pi (l\Delta t\nu-q_{l}\Delta F\tau)}$. 

According to \eqref{eq:FSTCRV}, each element of the FSTCRV comprises a superposition of channels of all paths, resulting in the non-sparsity of FST-domain channels and posing a great challenge for channel estimation and prediction with limited pilot overhead.  
Considering the high correlation caused by the limited scattering property of massive MIMO-OFDM channels \cite{wang2018survey}, we can leverage the eigenvector matrix to decouple the FSTCRV into a sparsity-domain one. 
The conventional eigenvector matrices for the frequency-domain CRV of OFDM system, the space-domain CRV of uniform linear array (ULA), and the time-domain CRV with uniform sampling can be expressed by DFT matrix \cite{barbotin12estimation,adhikary2013joint,shi2021compressive}. 
The delay-angle-Doppler-domain channel response vector (DADCRV) on the $l$-th subband based on DFT $ \bar{\mathbf{h}}_{l}\in\mathbb{C}^{NMK} $ is denoted as
\begin{align}
	\label{eq:DADCRV DFT}
	&\bar{\mathbf{h}}_{l} = \left(\mathbf{F}^{H}_{N}\otimes \bar{\mathbf{F}}^{H}_{M_{\text{v}}}\otimes\bar{\mathbf{F}}^{H}_{M_{\text{v}}}\otimes\bar{\mathbf{F}}_{K}\right)\mathbf{g}_{l} \\
	&= \hspace{-2mm}\int_{0}^{\tau_{\text{max}}}\hspace{-2mm}\int_{0}^{\pi}\hspace{-2mm}\int_{0}^{\pi}\hspace{-2mm}\int_{\nu_{\text{min}}}^{\nu_{\text{max}}}\hspace{-2mm}a(\tau,\theta,\varphi,\nu)
	\bar{\mathbf{w}}(\tau,\theta,\varphi,\tau)
	\psi_{l}(\tau,\nu)
	\text{d}\nu\text{d}\varphi\text{d}\theta\text{d}\tau,\nonumber
\end{align}
where $ \mathbf{F}_{N} $ and $ \bar{\mathbf{F}}_{N} $ are the DFT and phase-shift DFT matrices, which are defined by $ [\mathbf{F}_{N}]_{i,j}\triangleq\frac{1}{\sqrt{N}}e^{-\bar{\jmath}2\pi\frac{ij}{N}} $ and $ [\bar{\mathbf{F}}_{N}]_{i,j} \triangleq \frac{1}{\sqrt{N}}e^{-\bar{\jmath}2\pi\frac{i(j-M/2)}{N}} $, respectively. 
The vector $ \bar{\mathbf{w}}(\tau,\theta,\varphi,\tau) $ is defined by
\begin{align}
	\bar{\mathbf{w}}(\tau,\theta,\varphi,\tau)= \bar{\mathbf{b}}(\tau)\otimes \bar{\mathbf{c}}_{\text{v}}(\theta)\otimes\bar{\mathbf{c}}_{\text{h}}(\varphi;\theta)\otimes \bar{\mathbf{d}}(\nu),
\end{align}
where $ \bar{\mathbf{b}}(\tau)=\mathbf{F}^{H}_{N}\mathbf{b}(\tau) $, 
$ \bar{\mathbf{c}}_{\text{v}}(\theta)=\bar{\mathbf{F}}^{H}_{M_{\text{v}}}\mathbf{c}_{\text{v}}(\theta) $, 
$ \bar{\mathbf{c}}_{\text{h}}(\varphi;\theta)=\bar{\mathbf{F}}^{H}_{M_{\text{v}}}\mathbf{c}_{\text{h}}(\varphi;\theta) $, 
$ \bar{\mathbf{d}}(\nu)=\bar{\mathbf{F}}_{K}\mathbf{d}(\nu) $. 

The expressions of $ \bar{\mathbf{b}}(\tau) $, $ \bar{\mathbf{c}}_{\text{v}}(\theta) $, $ \bar{\mathbf{c}}_{\text{h}}(\varphi;\theta) $, and $ \bar{\mathbf{d}}(\nu) $ are similar and obtained by applying the IDFT or DFT to the steering vectors. 
For instance, the $ n $-th element of $ \bar{\mathbf{b}}(\tau) $ is 
\begin{align}
	[\bar{\mathbf{b}}(\tau)]_{n} =f_{N}\left(\Delta f\tau-n/N\right),
\end{align}
where the function $ f_{N}(x) $ is defined as
\begin{align}
	\textstyle f_{N}(x)\triangleq\frac{1}{\sqrt{N}}e^{-\bar{\jmath}\pi(N-1)x}\frac{\sin\left(\pi Nx\right)}{\sin\left(\pi x\right)}.
\end{align}
Based on the above simplification, the function $ [\bar{\mathbf{b}}(\tau)]_{n} $ can be regarded as a Sinc sampling function \cite{wu2021channel} in the delay domain and its peak point is $ (\bar{\tau}_{n}^{N},\sqrt{N}) $, where $ \bar{\tau}_{n}^{N} \triangleq \frac{n}{N\Delta f} $. Note that we define $ \bar{\tau}_{n}^{N}$ with $ n\in\mathbb{Z} $ and $ \bar{\tau}_{n^{\prime}}$ with $ n^{\prime}\notin\mathbb{Z} $ as the on-grid delay and off-grid delay, respectively. 
Similarly, the functions $ [\bar{\mathbf{c}}_{\text{v}}(\theta)]_{m_{\text{v}}} $, $[\bar{\mathbf{c}}_{\text{h}}(\varphi;\theta)]_{m_{\text{h}}}$, and $ [\bar{\mathbf{d}}(\nu)]_{k} $ can also be regarded as the Sinc sampling functions in the elevation angle, azimuth angle, and Doppler domains, whose sampling grids are defined by $ \bar{\theta}_{m_{\text{v}}}^{M_{\text{v}}}\triangleq\arccos( \frac{2m_{\text{v}}}{M_{\text{v}}}-1 ) $, $ \bar{\varphi}_{m_{\text{h}}}^{M_{\text{h}}}\triangleq \arccos(( \frac{2m_{\text{h}}}{M_{\text{h}}}-1 )\frac{1}{\sin\theta}) $, and $ \bar{\nu}_{k}^{K}\triangleq\frac{k-K/2}{K\Delta T} $.

Then, we analyze the asymptotical property of sampling functions based on DFT, as summarized in Lemma 1.

\textit{Lemma 1:} In the infinite case, the sampling functions are asymptotically equal to the Kronecker delta functions.

\textit{Proof:} Since the sampling functions are the special cases of the function $ f_{N}(x) $, we analyze $ f_{N}(x) $ firstly. When $ N $ tends to infinity, $ \lim_{N\rightarrow +\infty}f_{N}(x) $ is not equal to $ 0 $ if and only if $ x $ is an integer. 
Taking $ [\bar{\mathbf{b}}(\tau)]_{n} $ as an example, $ -1<\Delta f\tau-\frac{n}{N}<1 $ because $ 0\le n <N $ and $ 0\le\tau<\frac{1}{\Delta f} $. Therefore, $ [\bar{\mathbf{b}}(\tau)]_{n} $ satisfies the following property 
\begin{align}
	\lim_{N\rightarrow +\infty}\frac{1}{\sqrt{N}}[\bar{\mathbf{b}}(\tau)]_{n}&=
	\left\lbrace
	\begin{array}{ll}
		1, \hspace{-1mm}&\tau=\bar{\tau}_{n}^{N},\\
		0,\hspace{-1mm}&\text{otherwise}.
	\end{array}
	\right.
	\hspace{-1mm}=\delta[\tau-\bar{\tau}_{n}^{N}].
\end{align}
The derivations of other sampling functions follow a similar pattern. This completes the proof.

Inspired by Lemma 1, in the infinite case, the DADCR based on DFT is equivalent to sampling at fixed grids in the DAD domain by substituting the asymptotical expressions of sampling functions into \eqref{eq:DADCRV DFT}. In addition, the intervals between the adjacent sampling grids based on DFT become narrower as the dimension of DFT increases, indicating that the sampling grids can match the true values exactly under infinite conditions without any energy leakage. 

\subsection{Off-grid Basis for Delay-Angle-Doppler-Domain Channel}\label{subsec:off-grid basis}
In practical scenarios, the infinite condition may not be satisfied feasibly for the following reasons:
\begin{itemize}
	\item In the FHS scenarios, the fullband is divided into $ L $ subbands with a definite number of subcarriers. 
	\item BSs are often configured with a limited number of antennas due to cost considerations.
	\item In moderate or high-mobility scenarios, the number of time-domain channel samples is finite over the stationary time.
\end{itemize}
Thus, the sampling intervals based on DFT may not be small enough in the definite case, potentially causing the sampling grids to misalign with the true values. This off-grid case can lead to the energy leakage problem \cite{lian2019exploiting}.

\begin{figure}[!tb]
	\centering
	\subfigure[]{
		\includegraphics[width=0.5\linewidth]{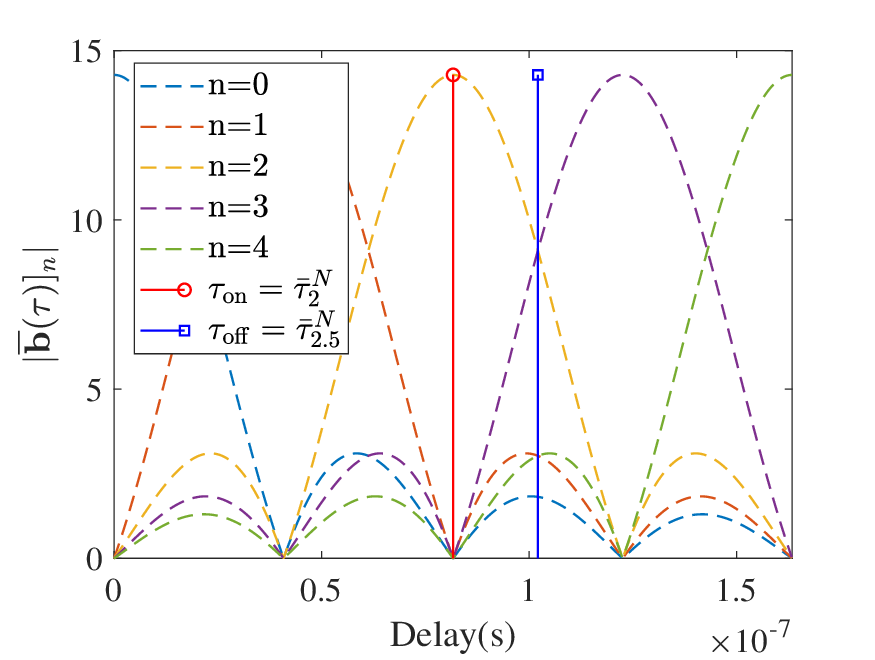}}
	\hspace{-5mm}
	\subfigure[]{
		\includegraphics[width=0.5\linewidth]{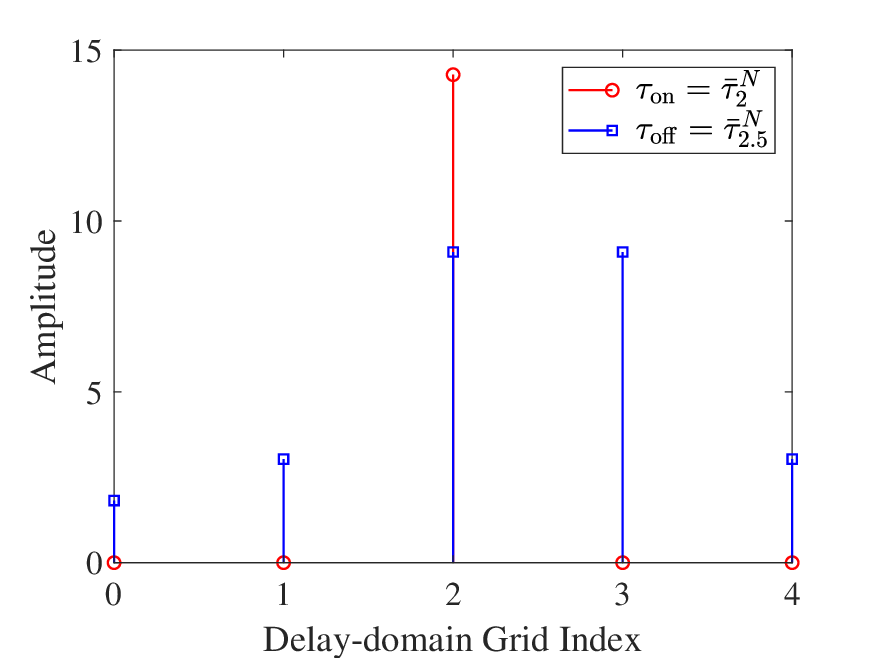}}
	\caption{Illustration of the energy leakage problem based on DFT with $ N=204 $ and $ \Delta f = 120 $ kHz: (a) The function curves of $ \lvert[\bar{\mathbf{b}}(\tau)]_{n}\rvert $ with the on-grid delay $ \tau_{\text{on}} = \bar{\tau}_{2}^{N} $ and the off-grid delay $ \tau_{\text{on}} = \bar{\tau}_{2.5}^{N} $. (b) The corresponding amplitudes of delay-domain grids in the on-grid and off-grid case.}
	\label{fig:off grid and energy leakage}
\end{figure}

To better illustrate the energy leakage problem based on DFT in the definite case, we take the delay-domain channel as an example. As shown in Fig. \ref{fig:off grid and energy leakage}(a), the function curves of $ \lvert[\bar{\mathbf{b}}(\tau)]_{n}\rvert $ with different indices $ n $ are denoted by the dotted lines. 
The on-grid delay $ \tau_{\text{on}} = \bar{\tau}_{2}^{N} $ and the off-grid delay $ \tau_{\text{off}} = \bar{\tau}_{2.5}^{N} $ are denoted by the red line and the blue line, respectively.
For the on-grid case, only the amplitude of the delay-domain grid corresponding to the on-grid delay $ \tau_{\text{on}} $ is nonzero, as validated by the red lines in Fig. \ref{fig:off grid and energy leakage}(b).
In contrast, for the off-grid delay $ \tau_{\text{off}} $, there exists energy leaking into all delay-domain grids, as shown by the blue lines in Fig. \ref{fig:off grid and energy leakage}(b). Similar analyses in the angle domain and Doppler domain are omitted owing to space limitations.
Based on the above analyses, the energy leakage problem can weaken sparsity and enrich multipath components of the DAD-domain grids, which poses significant challenges for channel estimation and prediction.

We assume that the scattering environment comprises $ P $ paths in total. The actual delays, angles, and Doppler shifts are represented by $ \left\lbrace \tau_{p},\theta_{p},\varphi_{p},\nu_{p} \right\rbrace_{p=0}^{P-1} $. 
The DAD-domain sampling grids are 
$ \lbrace \bar{\tau}_{n} \rbrace_{n=0}^{\tilde{N}-1} $, 
$ \lbrace\bar{\theta}_{m_{\text{v}}} \rbrace_{m_{\text{v}}=0}^{\tilde{M}_{\text{v}}-1} $, 
$ \lbrace\bar{\varphi}_{m_{\text{h}}} \rbrace_{m_{\text{h}}=0}^{\tilde{M}_{\text{h}}-1} $, and 
$ \lbrace \bar{\nu}_{k} \rbrace_{k=0}^{\tilde{K}-1} $, 
where $ \tilde{N} $, $ \tilde{M}_{\text{v}} $, $ \tilde{M}_{\text{h}} $, and $ \tilde{K} $ are the numbers of sampling grids covering the DAD domain uniformly. 
For simplicity, we omit the superscripts of the DAD-domain sampling grids and define $\tilde{M}=\tilde{M}_{\text{v}}\tilde{M}_{\text{h}}$.
To fulfill the demand for time-domain channel prediction, we set $ \tilde{K}=S_{\nu}K $ to refine the Doppler-domain parameter estimation, where $ S_{\nu} $ is the corresponding oversampling factor. 
In the delay-angle domain, we opt to set $ \tilde{N}=N $, $ \tilde{M}_{\text{v}}=M_{\text{v}} $, and $ \tilde{M}_{\text{h}}=M_{\text{h}} $ to facilitate the complexity reduction in Section \ref{sec low-complexity hyper-parameter}.

As mentioned above, it is impractical to assume the true values are located exactly on the fixed DAD-domain grids. 
Therefore, we introduce the off-grid model to narrow the gaps between the true values and sampling grids, which is given by
\begin{align}
	x_{p} = \bar{x}_{n_{p}} \hspace{-0.5mm}+\hspace{-0.5mm} y_{n_{p}}, y_{n_{p}}\hspace{-0.5mm}\in\hspace{-0.5mm}\left[ \frac{\bar{x}_{n_{p}-1}-\bar{x}_{n_{p}}}{2},\frac{\bar{x}_{n_{p}+1}-\bar{x}_{n_{p}}}{2} \right),
\end{align}
where $x\in\lbrace\tau,\theta,\varphi,\nu\rbrace$ and $y\in\lbrace \alpha,\beta,\gamma,\eta \rbrace$. $\bar{x}_{n_{p}}$ is the nearest sampling grid to $x_{p}$ and $y_{n_{p}}$ corresponds to the off-grid gap. By permuting the steering vectors as matrices in the index order of sampling grids, we can obtain the corresponding off-grid matrices $\mathbf{B}(\boldsymbol{\alpha})$, $\mathbf{C}_{\text{v}}(\boldsymbol{\beta})$, $\mathbf{C}_{\text{h}}(\boldsymbol{\gamma})$, and $\mathbf{D}(\boldsymbol{\eta})$ with off-grid hyper-parameter vector $\boldsymbol{\alpha}$, $\boldsymbol{\beta}$, $\boldsymbol{\gamma}$, and $\boldsymbol{\eta}$, which are defined as 
$[\mathbf{B}(\boldsymbol{\alpha})]_{:,n}\triangleq \mathbf{b}(\bar{\tau}_{n}+\alpha_{n})$, 
$ [\mathbf{C}_{\text{v}}(\boldsymbol{\beta})]_{:,m_{\text{v}}} \triangleq \mathbf{c}_{\text{v}}(\bar{\theta}_{m_{\text{v}}}+\beta_{m_{\text{v}}}) $, 
$ [\mathbf{C}_{\text{h}}(\boldsymbol{\gamma})]_{:,m_{\text{h}}} \triangleq \mathbf{c}_{\text{h}}(\bar{\varphi}_{m_{\text{h}}}+\gamma_{m_{\text{h}}}) $, and $ [\mathbf{D}(\boldsymbol{\eta})]_{:,k} \triangleq \mathbf{d}(\bar{\nu}_{k}+\eta_{k}) $. 

\textit{Remark 1: }The DFT transform matrix is one special case of the above off-grid matrices when true values align with uniform sampling grids and the number of sampling grids is equal to that of observations. 

Therefore, the transform between the FSTCRV $ \mathbf{g}_{l} $ and the DADCRV based on accurate off-grid matrix $ \tilde{\mathbf{h}}_{l} $ is given by
\begin{align}
	\label{eq:FST DAD off-grid}
	\mathbf{g}_{l} = 
	\tilde{\mathbf{W}}(\boldsymbol{\omega})\tilde{\mathbf{h}}_{l}, 
\end{align} 
where $ \boldsymbol{\omega} \triangleq [\boldsymbol{\alpha}^{T},\boldsymbol{\beta}^{T},\boldsymbol{\gamma}^{T},\boldsymbol{\eta}^{T}]^{T} $. $ \tilde{\mathbf{W}}(\boldsymbol{\omega})\triangleq\mathbf{B}(\boldsymbol{\alpha})\otimes \mathbf{C}(\boldsymbol{\beta},\boldsymbol{\gamma})\otimes \mathbf{D}(\boldsymbol{\eta}) $ denotes the accurate DAD-domain off-grid matrix and $\mathbf{C}(\boldsymbol{\beta},\boldsymbol{\gamma})\triangleq \mathbf{C}_{\text{v}}(\boldsymbol{\beta})\otimes \mathbf{C}_{\text{h}}(\boldsymbol{\gamma})$. 

\textit{Remark 2: }The channel model \eqref{eq:FST DAD off-grid} can be degraded into the delay-angle-domain channel model in \cite{wan2024two} by substituting the Doppler-domain off-grid matrix $ \mathbf{D}(\boldsymbol{\eta}) $ into the identity matrix $ \mathbf{I}_{K} $, and further degraded into the angle-domain channel model in \cite{lian2019exploiting} by substituting the delay-domain off-grid matrix $ \mathbf{B}(\boldsymbol{\alpha}) $ into the identity matrix $ \mathbf{I}_{N} $.

To trade off the tractability and fidelity, the accurate DAD-domain off-grid matrix is approximated by Taylor series expansion as 
\begin{align}
	\mathbf{W}(\boldsymbol{\omega}) = &\mathbf{W} + \dot{\mathbf{W}}_{\alpha}\text{diag}\lbrace\mathbf{R}_{\alpha}\boldsymbol{\alpha}\rbrace
	+ \dot{\mathbf{W}}_{\beta}\text{diag}\lbrace\mathbf{R}_{\beta}\boldsymbol{\beta}\rbrace \nonumber\\
	&+ \dot{\mathbf{W}}_{\gamma}\text{diag}\lbrace\mathbf{R}_{\gamma}\boldsymbol{\gamma}\rbrace
	+ \dot{\mathbf{W}}_{\eta}\text{diag}\lbrace\mathbf{R}_{\eta}\boldsymbol{\eta}\rbrace,
\end{align}
where $\mathbf{W}\triangleq\tilde{\mathbf{W}}(\mathbf{0}_{\tilde{N}\tilde{M}\tilde{K}})$ denotes the transform matrix without off-grid hyper-parameters. 
$ \dot{\mathbf{W}}_{x} $ is the first-order partial derivation of W concerning $x\in\lbrace\tau,\theta,\varphi,\nu\rbrace$.
$ \mathbf{R}_{\alpha}\triangleq \mathbf{I}_{\tilde{N}}\otimes \mathbf{1}_{\tilde{M}_{\text{v}}\tilde{M}_{\text{h}}\tilde{K}} $,
$ \mathbf{R}_{\beta}\triangleq \mathbf{1}_{\tilde{N}}\otimes \mathbf{I}_{\tilde{M}_{\text{v}}}\otimes \mathbf{1}_{\tilde{M}_{\text{h}}\tilde{K}} $, 
$\mathbf{R}_{\gamma}\triangleq \mathbf{1}_{\tilde{N}\tilde{M}_{\text{v}}}\otimes \mathbf{I}_{\tilde{M}_{\text{h}}}\otimes \mathbf{1}_{\tilde{K}}$, 
and 
$ \mathbf{R}_{\eta}\triangleq \mathbf{1}_{\tilde{N}\tilde{M}_{\text{v}}\tilde{M}_{\text{h}}}\otimes \mathbf{I}_{\tilde{K}} $.

\section{Problem Formulation} 
\label{sec3}
In this section, we first analyze the statistical characteristics of subbands and formulate the MS JCEP problem as a generalized MMV problem. 
Subsequently, we develop a probability model and formulate the corresponding maximum a-posterior (MAP) estimation problem.

\subsection{JCEP Problem Formulation}
\begin{figure}[!tb]
	\includegraphics[width=1\linewidth]{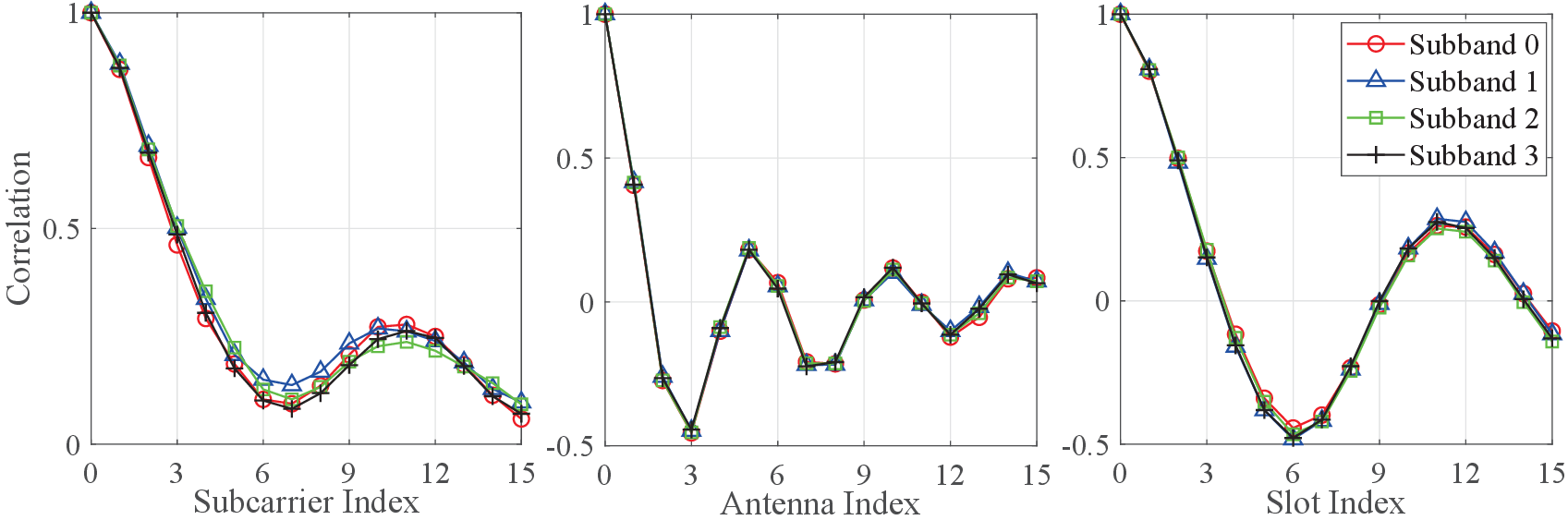}
	\caption{The frequency, spatial, and temporal correlation of different subbands.}
	\label{fig:FST corr}
\end{figure}

The FHS mode introduces the concept of subbands, which motivates us to explore the MS channel statistical characteristics for higher CSI accuracy.
The frequency, spatial, and temporal correlation of different subbands is shown in Fig.~\ref{fig:FST corr}, which suggests that they share a common correlation. The above observation will be included in Proposition 1. 

\textit{Proposition 1: }The statistical characteristics of the subband channel  $ \text{E}[ \mathbf{g}_{l}\left(\mathbf{g}_{l}\right)^{H} ] $ are independent of the subband index $l$.

\textit{Proof: }The correlation matrix of the FSTCRV on the $ l $-th subband is given by 
	\begin{align}\label{eq:FST corr}
		&\text{E}\left[ \mathbf{g}_{l}\left(\mathbf{g}_{l}\right)^{H} \right]
		\overset{\text{(a)}}{=} 
		\int_{0}^{\tau_{\text{max}}}\hspace{-2mm}\int_{0}^{\pi}\hspace{-2mm}\int_{0}^{\pi}\hspace{-2mm}\int_{\nu_{\text{min}}}^{\nu_{\text{max}}}
		A(\tau,\theta,\varphi,\nu) \nonumber\\
		&\qquad\quad\cdot\mathbf{w}(\tau,\theta,\varphi,\nu)\left(\mathbf{w}(\tau,\theta,\varphi,\nu)\right)^{H}
		\text{d}\nu\text{d}\varphi\text{d}\theta\text{d}\tau,
	\end{align}
where equation (a) is obtained by plugging \eqref{eq:E complex gain} in $ \text{E}[ \mathbf{g}_{l}\left(\mathbf{g}_{l}\right)^{H} ] $ and simplifying. This completes the proof. 

Following the analysis of the statistical property of a single subband channel, we delve into the correlation between different subband channels, as summarized in Proposition 2.

\textit{Proposition 2:} The same channel elements of different subbands are independent of each other in rich scattering scenarios. 

\textit{Proof:} The correlation between the same channel elements of different subbands is expressed as
\begin{align}
	\text{E}[g_{nmkl}g_{nmkl^{\prime}}^{*}] &= \int_{0}^{\tau_{\text{max}}}\hspace{-2mm}\int_{0}^{\pi}\hspace{-2mm}\int_{0}^{\pi}\hspace{-2mm}\int_{\nu_{\text{min}}}^{\nu_{\text{max}}}\hspace{-2mm}
	A(\tau,\theta,\varphi,\nu) \nonumber\\
	&\qquad\cdot\psi_{l}(\tau,\nu)\psi_{l^{\prime}}^{*}(\tau,\nu)
	\text{d}\nu\text{d}\varphi\text{d}\theta\text{d}\tau. 
\end{align}
Notably, the correlation $\text{E}[g_{nmkl}g_{nmkl^{\prime}}^{*}]$ equals zero in rich scattering scenarios, as per the central limit theorem \cite{liu2018massive}. This concludes the proof.

Proposition 1 and Proposition 2 reveal that the same channel elements of different subbands are i.i.d., allowing us to formulate the MS JCEP problem as a generalized MMV problem. 
We can operate CSI acquisition of different subbands independently while exploiting the common channel statistical characteristics of different subbands, potentially enhancing JCEP performance. 
We assume that the UE transmits identical SRS sequences across different subbands \cite{38.211}. 
Therefore, the received signal vectors of different subbands can be consolidated into a single matrix $ \mathbf{Y} = [ \mathbf{y}_{1},\ldots,\mathbf{y}_{L} ] $, i.e., 
\begin{align}
	\label{eq:received signal model2}
	\mathbf{Y} &= \mathbf{S}\mathbf{W}(\boldsymbol{\omega}) \mathbf{H} + \mathbf{Z},
\end{align}
where $\mathbf{H}$ is the delay-angle-Doppler-domain channel response matrix (DADCRM) based on the approximate off-grid matrix, whose the $ (\tilde{n}\tilde{M}\tilde{K} + \tilde{m}\tilde{K} + \tilde{k},l) $-th element is $ h_{\tilde{n}\tilde{m}\tilde{k}l} $. 
$ \mathbf{G}\triangleq\mathbf{W}(\boldsymbol{\omega}) \mathbf{H} $ is the frequency-space-time-domain channel response matrix (FSTCRM).
It is worth noting that the transform matrix $ \mathbf{W}(\boldsymbol{\omega}) $ is shared among DADCRVs across different subbands due to Proposition 1. 

\textit{Remark 3: }Specifically, when the FHS mode is disabled, i.e., $L=1$, the single-subband (SS) JCEP problem can be modeled as a single measurement vector (SMV) problem, which is a particular case of the MMV problem.

\subsection{Probability Model}
\label{sec MAP}
From the received signal model \eqref{eq:received signal model2}, we present the channel transfer function to characterize the signal transmission and the prior statistical channel model to capture the sparsity of the DAD-domain channel. After the probability modeling, the MMV problem can be formulated as one MAP estimation problem. 

\subsubsection{Channel Transfer Function}
Based on the received signal model \eqref{eq:received signal model2}, the channel transfer function is expressed as
\begin{align}
	p(\mathbf{Y}\vert \mathbf{H}; \boldsymbol{\omega}) = p\left(\mathbf{Y}\vert\mathbf{G}\right)
	p\left(\mathbf{G}\vert\mathbf{H};\boldsymbol{\omega}\right).
\end{align}
Because of the independence between noise elements, the likelihood function $ p(\mathbf{Y}\vert\mathbf{G}) $ can be factorized as 
\begin{align}
	p(\mathbf{Y}\vert\mathbf{G})&=\prod_{n,m,k,l}p( y_{nmkl}\vert g_{nmkl} ) \nonumber\\
	&=\prod_{n,m,k,l}\mathcal{CN}(y_{nmkl};s_{nmk}g_{nmkl},\sigma_{\text{z}}).
\end{align}
Considering the large-scale condition, we factorize the conditional function $ p(\mathbf{G}\vert\mathbf{H};\boldsymbol{\omega}) $ as 
\begin{align}
	p(\mathbf{G}\vert\mathbf{H};\boldsymbol{\omega})&=\prod_{n,m,k,l}p( g_{nmkl}\vert\mathbf{h}_{l};\boldsymbol{\omega} )\nonumber\\
	&=\prod_{n,m,k,l}\delta( g_{nmkl}- 
	(\underline{\mathbf{w}}_{nmk}(\boldsymbol{\omega}))^{T}
	\mathbf{h}_{l} ),
\end{align}
where $\mathbf{h}_{l}$ is the $l$-th column vector of $\mathbf{H}$ and
$\underline{\mathbf{w}}_{nmk}(\boldsymbol{\omega}) $ is the $(nMK+mK+k)$-th row vector of $ \mathbf{W}(\boldsymbol{\omega}) $. 
To facilitate hyper-parameter learning, we approximate the factor function $ p( g_{nmkl}\vert\mathbf{h}_{l};\boldsymbol{\omega} ) $ as a complex Gaussian distribution with variance approaching zero, i.e., $\lim_{\epsilon\rightarrow 0}\mathcal{CN}( g_{nmkl};(\underline{\mathbf{w}}_{nmk}(\boldsymbol{\omega}))^{T}\mathbf{h}_{l},\epsilon)$ \cite{ghatak2004quantum}. 

\subsubsection{Prior Statistical Channel Model}
After the off-grid modeling of DAD-domain channels, the energy leakage problem is efficiently resolved. 
Compared with elements of DADCRM with DFT basis, those with off-grid basis exhibit better independence and sparsity due to fewer multipath components. 
To characterize independence and sparsity, we introduce independently and non-identically distributed (i.n.d.) conditional Bernoulli-Gaussian (BG) distribution \cite{Liu18Massive} to the elements of $\mathbf{H}$ as
\begin{align}
	&p(\mathbf{H}\vert \boldsymbol{\xi};\boldsymbol{\sigma}) 
	=\hspace{-2mm} \prod_{\tilde{n},\tilde{m},\tilde{k},l} \hspace{-2mm}
	p ( h_{\tilde{n}\tilde{m}\tilde{k}l}\vert\xi_{\tilde{n}\tilde{m}\tilde{k}};\sigma_{\tilde{n}\tilde{m}\tilde{k}} ) \\
	&=\hspace{-2mm} \prod_{\tilde{n},\tilde{m},\tilde{k},l} \hspace{-2mm}
	\left(\delta[\xi_{\tilde{n}\tilde{m}\tilde{k}}\hspace{-1mm}-\hspace{-1mm}1]\mathcal{CN}\left( h_{\tilde{n}\tilde{m}\tilde{k}l};0,\sigma_{\tilde{n}\tilde{m}\tilde{k}} \right)  \hspace{-1mm}+\hspace{-1mm} \delta[\xi_{\tilde{n}\tilde{m}\tilde{k}}]\delta(h_{\tilde{n}\tilde{m}\tilde{k}l})\right), \nonumber
\end{align}
where $ \xi_{\tilde{n}\tilde{m}\tilde{k}} $ and $ \sigma_{\tilde{n}\tilde{m}\tilde{k}} $ are the $(\tilde{n}\tilde{M}\tilde{K} + \tilde{m}\tilde{K} + \tilde{k})$-th elements of the state indicator vector $ \boldsymbol{\xi}\in\mathbb{B}^{\tilde{N}\tilde{M}\tilde{K}} $ and the prior variance vector $\boldsymbol{\sigma}\in\mathbb{R}^{\tilde{N}\tilde{M}\tilde{K}}$.
Note that the DAD-domain channel elements of different subbands can be processed independently and share the same state indicator and prior variance.
The probability density function (PDF) of $ \boldsymbol{\xi} $ follows the Bernoulli distribution as \cite{zhu2022ofdm}
\begin{align}
	p(\boldsymbol{\xi};\rho) = \prod_{\tilde{n},\tilde{m},\tilde{k}} \left(\rho\delta[\xi_{\tilde{n}\tilde{m}\tilde{k}}-1] + (1-\rho)\delta[\xi_{\tilde{n}\tilde{m}\tilde{k}}]\right),
\end{align} 
where $ \rho $ is the prior active probability. 

\subsubsection{Maximum A-posterior Estimation Problem}

\begin{figure}[!tb]
	\centering
	\includegraphics[width=1\linewidth]{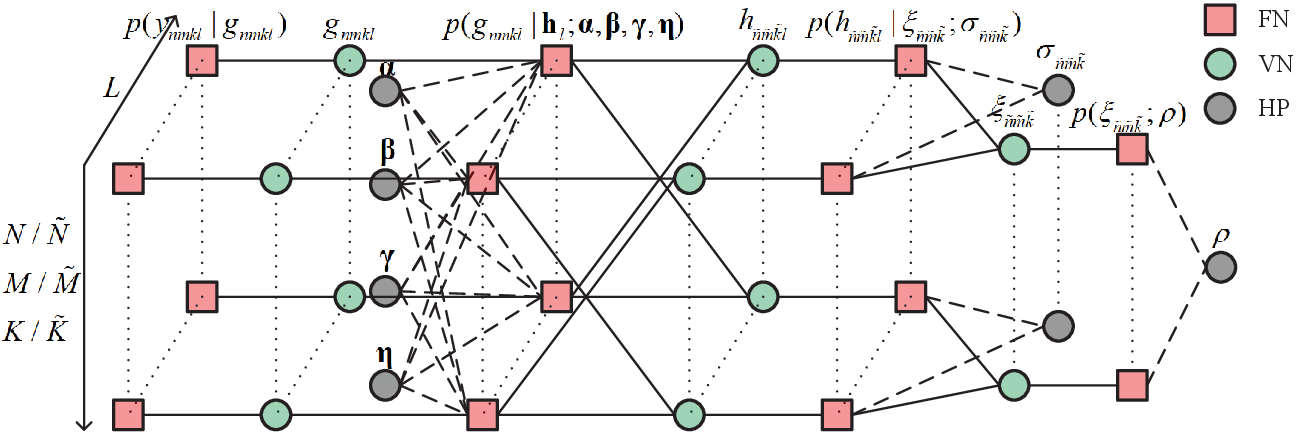}
	\caption{Factor graph of the factorized result of joint PDF. The red squares, green circles, and grey circles denote factor nodes, variable nodes, and hyper-parameters, respectively.}
	\label{fig:factor graph}
\end{figure}

The JCEP problem based on the received signal model \eqref{eq:received signal model2} is to estimate the DADCRM $ \mathbf{H} $ based on the received signal matrix $ \mathbf{Y} $ and pilot matrix $ \mathbf{S} $. Combined with the above probability model, the optimal MAP channel estimator is expressed as
\begin{align}\label{eq:MAP estimator}
	&\hat{\mathbf{H}}\hspace{-1mm}= \hspace{-1mm}\underset{\mathbf{H}}{\text{arg max}}\: 
	\sum_{\boldsymbol{\xi}} \int
	p\left( \mathbf{G},\mathbf{H},\boldsymbol{\xi}\vert \mathbf{Y};\boldsymbol{\omega},\boldsymbol{\sigma},\rho \right)
	\text{d}\mathbf{G} \nonumber\\
	&\hspace{-1mm}=\hspace{-1mm}\underset{\mathbf{H}}{\text{arg max}}\:\frac{1}{Z(\mathbf{Y})}\sum_{\boldsymbol{\xi}} \int
	p\left(
	\mathbf{Y},\mathbf{G},\mathbf{H},\boldsymbol{\xi};\boldsymbol{\omega},\boldsymbol{\sigma},\rho
	\right)
	\text{d}\mathbf{G} \\
	&\hspace{-1mm}=\hspace{-1mm}\underset{\mathbf{H}}{\text{arg max}}\:\sum_{\boldsymbol{\xi}} \int
	p\left(\mathbf{Y}\vert\mathbf{G}\right)
	p\left(\mathbf{G}\vert\mathbf{H};\boldsymbol{\omega}\right)
	p\left(\mathbf{H}\vert\boldsymbol{\xi};\boldsymbol{\sigma}\right)
	p\left(\boldsymbol{\xi};\rho\right)
	\text{d}\mathbf{G}, \nonumber
\end{align}
where $ Z\left( \mathbf{Y} \right) $ denotes the normalization constant and 
$ p\left(
\mathbf{Y},\mathbf{G},\mathbf{H},\boldsymbol{\xi};\boldsymbol{\omega},\boldsymbol{\sigma},\rho
\right) $
is the corresponding joint PDF. 
Given the above factorized results of joint PDF, we draw the corresponding factor graph \cite{kschischang2001factor} in Fig. \ref{fig:factor graph} to facilitate the visualization of the probability model.

From \eqref{eq:MAP estimator}, the optimal MAP channel estimator necessitates the complicated computation of marginals of joint PDF involving multi-dimensional integrals and summations, particularly in large-scale systems.

\section{Joint Channel Estimation and Prediction}
\label{sec4}
In this section, we start from the perspective of variational Bayesian inference (VBI) \cite{Yedidia05Constructing} and derive the Off-Grid-MS HMP algorithm under the received signal \eqref{eq:received signal model2} to solve the JCEP problem with FHS. Aiming at the lack of prior CSI in practical scenarios, we utilize the proposed algorithm to adaptively learn the hyper-parameters and reduce the computational complexity of off-grid hyper-parameters by leveraging the approximations of the off-grid matrices.

\subsection{Bethe Free Energy Minimization}

To reduce the complexity of statistical inference, we investigate the VBI, which approximates posterior PDF by introducing corresponding beliefs \cite{Yedidia05Constructing}.  
According to the optimization problem \eqref{eq:MAP estimator}, we introduce the belief $b(\mathbf{G},\mathbf{H},\boldsymbol{\xi},\boldsymbol{\omega},\boldsymbol{\sigma},\rho)$ to approximate joint posterior PDF. 
The belief can be obtained by minimizing Kullback-Leibler (KL) divergence characterizing the similarity of belief and actual PDF, which is written as 
\begin{align}\label{eq:KL divergence}
    &\hat{b}(\mathbf{G},\mathbf{H},\boldsymbol{\xi},\boldsymbol{\omega},\boldsymbol{\sigma},\rho)\nonumber \\
    &=\hspace{-3mm}\underset{b(\mathbf{G},\mathbf{H},\boldsymbol{\xi},\boldsymbol{\omega},\boldsymbol{\sigma},\rho)\in\mathcal{Q}}{\text{arg min}}\hspace{-3mm}\text{D}[b(\mathbf{G},\mathbf{H},\boldsymbol{\xi},\boldsymbol{\omega},\boldsymbol{\sigma},\rho)\Vert p(\mathbf{G},\mathbf{H},\boldsymbol{\xi}\vert \mathbf{Y};\boldsymbol{\omega},\boldsymbol{\sigma},\rho)]\nonumber\\
    &=\underset{b(\mathbf{G},\mathbf{H},\boldsymbol{\xi},\boldsymbol{\omega},\boldsymbol{\sigma},\rho)\in\mathcal{Q}}{\text{arg min}}{\mathcal{F}_{\text{V}}}+\ln  Z(\mathbf{Y}),
\end{align}
where $\mathcal{Q}$ denotes the constrained set of belief $b(\mathbf{G},\mathbf{H},\boldsymbol{\xi},\boldsymbol{\omega},\boldsymbol{\sigma},\rho)$.
$\mathcal{F}_{\text{V}}$ represents variational free energy (VFE), which is denoted as
\begin{align}
	\mathcal{F}_{\text{V}}&\triangleq
	\text{D}[b(\mathbf{G},\mathbf{H},\boldsymbol{\xi},\boldsymbol{\omega},\boldsymbol{\sigma},\rho)\Vert p(\mathbf{G},\mathbf{H},\boldsymbol{\xi}, \mathbf{Y};\boldsymbol{\omega},\boldsymbol{\sigma},\rho)].
\end{align}
Hence, the MAP problem is transformed into the VFE minimization problem. 
 
The complicated dependency between random variables renders the above optimal problem intractable, especially in large-scale systems. 
The design of constrained set $\mathcal{Q}$ plays an important role in balancing the complexity and accuracy of statistical inference. 
Therefore, the Bethe approximation method is proposed to design the constrained set of beliefs by analyzing the dependency between different variables to retain local dependency in the inference process \cite{Yedidia05Constructing,Zhang21Unifying}.
According to the factorized results of joint PDF in Section \ref{sec MAP}, we introduce auxiliary beliefs for factor functions and random variables, as shown in Table \ref{tab:belief}. 
\begin{table}[!t]
	\small
	\centering
	\caption{The Beliefs for Factor Functions and Random Variables}
	\label{tab:belief}
	\begin{tabular}{cc}
		\hline
		\textbf{Factor Function} & \textbf{Beliefs}  \\ \hline 
		$p( y_{nmkl}\vert g_{nmkl} )$ & $b^{g}_{nmkl}(g_{nmkl})$ \\  
		$p( g_{nmkl}\vert\mathbf{h}_{l};\boldsymbol{\omega} )$ & $b^{g\mathbf{h}\boldsymbol{\omega}}_{nmkl}(g_{nmkl},\mathbf{h}_{l}, \boldsymbol{\omega})$ \\ 
		$p ( h_{\tilde{n}\tilde{m}\tilde{k}l}\vert\xi_{\tilde{n}\tilde{m}\tilde{k}};\sigma_{\tilde{n}\tilde{m}\tilde{k}} )$  & $b^{h\xi\sigma}_{\tilde{n}\tilde{m}\tilde{k}l}(h_{\tilde{n}\tilde{m}\tilde{k}l},\xi_{\tilde{n}\tilde{m}\tilde{k}},\sigma_{\tilde{n}\tilde{m}\tilde{k}})$  \\ 
		$p( \xi_{\tilde{n}\tilde{m}\tilde{k}};\rho )$ & $b^{\xi\rho}_{\tilde{n}\tilde{m}\tilde{k}}(\xi_{\tilde{n}\tilde{m}\tilde{k}},\rho)$ \\ \hline
		 \textbf{Random Variables} &\textbf{Beliefs}\\  \hline
		$g_{nmkl}$ & $q^{g}_{nmkl}(g_{nmkl}) $\\ 
		$h_{\tilde{n}\tilde{m}\tilde{k}l}$ & $q^{h}_{\tilde{n}\tilde{m}\tilde{k}l}(h_{\tilde{n}\tilde{m}\tilde{k}l})$ \\ 
		$\xi_{\tilde{n}\tilde{m}\tilde{k}}$ & $q^{\xi}_{\tilde{n}\tilde{m}\tilde{k}}(\xi_{\tilde{n}\tilde{m}\tilde{k}})$ \\ \hline
	\end{tabular}
\end{table}
Based on the formulation rule of Bethe beliefs, we approximate joint PDF as 
\begin{align}\label{eq:Bethe belief}
    &b(\mathbf{G},\mathbf{H},\boldsymbol{\xi},\boldsymbol{\omega},\boldsymbol{\sigma},\rho)\\
    &
    =\frac{
	\prod_{n,m,k,l}
	b^{g}_{nmkl}b^{g\mathbf{h}\boldsymbol{\omega}}_{nmkl}
	\prod_{\tilde{n},\tilde{m},\tilde{k},l}
	b^{h\xi\sigma}_{\tilde{n}\tilde{m}\tilde{k}l}
	\prod_{\tilde{n},\tilde{m},\tilde{k}}
	b^{\xi\rho}_{\tilde{n}\tilde{m}\tilde{k}}}
	{
	\prod_{n,m,k,l}
	q^{g}_{nmkl}
	\prod_{\tilde{n},\tilde{m},\tilde{k},l}
	(q^{h}_{\tilde{n}\tilde{m}\tilde{k}l})^{NMK}
	\prod_{\tilde{n},\tilde{m},\tilde{k}}
	q^{\xi}_{\tilde{n}\tilde{m}\tilde{k}}}.\nonumber
\end{align}
Substituting \eqref{eq:Bethe belief} into the expression of VFE, we can obtain BFE as \eqref{eq:bethe free energy}.
 
\begin{figure*}[h]
	\begin{align}\label{eq:bethe free energy}
		{\mathcal{F}}_{\text{B}}=&\textstyle\sum_{n,m,k,l}\text{D}[b^{g}_{nmkl}\Vert p(y_{nmkl}\vert g_{nmkl})]
		+ \sum_{n,m,k,l}\text{D}
		[ b^{g\mathbf{h}\boldsymbol{\omega}}_{nmkl} \Vert p(g_{nmkl}\vert\mathbf{h}_{l};\boldsymbol{\omega} )]
		+
		\sum_{\tilde{n},\tilde{m},\tilde{k},l}\text{D}[b^{h\xi\sigma}_{\tilde{n}\tilde{m}\tilde{k}l} \Vert p( h_{\tilde{n}\tilde{m}\tilde{k}l}\vert\xi_{\tilde{n}\tilde{m}\tilde{k}};\sigma_{\tilde{n}\tilde{m}\tilde{k}} )]\nonumber\\
		&\textstyle+
		\sum_{\tilde{n},\tilde{m},\tilde{k}}\text{D}[b^{\xi\rho}_{\tilde{n}\tilde{m}\tilde{k}}\Vert p( \xi_{\tilde{n}\tilde{m}\tilde{k}};\rho )]
		+\sum_{n,m,k,l}\text{H}[q^{g}_{nmkl}]+\sum_{\tilde{n},\tilde{m},\tilde{k}}\text{H}[q^{\xi}_{\tilde{n}\tilde{m}\tilde{k}}]+
		\sum_{\tilde{n},\tilde{m},\tilde{k},l}NMK\text{H}[q^{h}_{\tilde{n}\tilde{m}\tilde{k}l}].
	\end{align}
	\hrulefill
\end{figure*} 

It is worth noting that if the beliefs of factor functions and random variables describe the PDF of the same variable, these beliefs satisfy marginal consistency constraints (MCCs). The VFE minimization problem is further transformed into the BFE minimization problem with MCCs. However, verifying MCCs entails high complexity, rendering such an optimization problem intractable. To address this issue, we aim to reduce complexity by redesigning constraints in Section \ref{sec4.2}.

\subsection{Redesign of Belief Constraints}\label{sec4.2}
Since hyper-parameters typically characterize the statistical properties of channel coefficients, we regard them as independent constants without prior information during the statistical inference. 
Firstly, we contemplate the factorization of beliefs concerning hyper-parameters. 
The prior active probability $\rho$ is the expectation of state indicator $\xi_{\tilde{n}\tilde{m}\tilde{k}}$, and thus exhibits a significantly slower variation rate compared with the instantaneous variable $\xi_{\tilde{n}\tilde{m}\tilde{k}}$. As such, the belief $b^{\xi\rho}_{\tilde{n}\tilde{m}\tilde{k}}$ can be factorized as
\begin{align}
	\label{eq:sec4 FC1}
	b^{\xi\rho}_{\tilde{n}\tilde{m}\tilde{k}}=g^{\xi}_{\tilde{n}\tilde{m}\tilde{k}}g^{\rho},\: g^{\rho} = \delta(\rho-\hat{\rho}),
\end{align}
where $g^{\xi}_{\tilde{n}\tilde{m}\tilde{k}}$ and $g^{\rho}$ are the beliefs factorized from the belief $ b^{\xi\rho}_{\tilde{n}\tilde{m}\tilde{k}} $ in regard to $\xi_{\tilde{n}\tilde{m}\tilde{k}}$ and $\rho$. 
$\hat{\rho}$ denotes the true value of $\rho$. 
As the variance of sparsity-domain channel coefficient $\xi_{\tilde{n}\tilde{m}\tilde{k}}h_{\tilde{n}\tilde{m}\tilde{k}l}$, the hyper-parameter $\sigma_{\tilde{n}\tilde{m}\tilde{k}}$ can also be treated as an independent constant during the process of JCEP, resulting in the factorization of belief $b^{h\xi\sigma}_{\tilde{n}\tilde{m}\tilde{k}l}$ as
\begin{align}
	\label{eq:sec4 FC2}
	b^{h\xi\sigma}_{\tilde{n}\tilde{m}\tilde{k}l}=b^{h\xi}_{\tilde{n}\tilde{m}\tilde{k}l}b^{\sigma}_{\tilde{n}\tilde{m}\tilde{k}},\:
	b^{\sigma}_{\tilde{n}\tilde{m}\tilde{k}} = \delta(\sigma_{\tilde{n}\tilde{m}\tilde{k}}-\hat{\sigma}_{\tilde{n}\tilde{m}\tilde{k}}),
\end{align}
where $b^{h\xi}_{\tilde{n}\tilde{m}\tilde{k}l}$ and $b^{\sigma}_{\tilde{n}\tilde{m}\tilde{k}}$ denote the beliefs factorized from the belief $b^{h\xi\sigma}_{\tilde{n}\tilde{m}\tilde{k}l}$ concerning $\xi_{\tilde{n}\tilde{m}\tilde{k}}h_{\tilde{n}\tilde{m}\tilde{k}l}$ and $\sigma_{\tilde{n}\tilde{m}\tilde{k}}$. 
$ \hat{\sigma}_{\tilde{n}\tilde{m}\tilde{k}} $ represents the true value of $ \sigma_{\tilde{n}\tilde{m}\tilde{k}} $.
Similarly, the belief $b^{g\mathbf{h}\boldsymbol{\omega}}_{nmkl}$ satisfies the following factorization constraints (FCs)
\begin{subequations}\label{eq:sec4 FC3}
	\begin{align}
		\label{eq:sec4 FC3-1}
		&b^{g\mathbf{h}\boldsymbol{\omega}}_{nmkl}=b^{g\mathbf{h}}_{nmkl}b^{\boldsymbol{\alpha}}b^{\boldsymbol{\beta}}b^{\boldsymbol{\gamma}}b^{\boldsymbol{\eta}},\\
		&b^{x} = \delta(x-\hat{x}),\quad x\in\lbrace \boldsymbol{\alpha},\boldsymbol{\beta},\boldsymbol{\gamma},\boldsymbol{\eta} \rbrace,
	\end{align}
\end{subequations}
where $\hat{x}$ denotes the true value of $x$.

Based on \eqref{eq:sec4 FC2}, we further define the following MCCs
\begin{subequations}\label{eq:sec4 MaCC1}
	\begin{align}
		b^{h}_{\tilde{n}\tilde{m}\tilde{k}l}&=\sum_{\xi_{\tilde{n}\tilde{m}\tilde{k}}\in\mathbb{B}}b^{h\xi}_{\tilde{n}\tilde{m}\tilde{k}l},\\
		b^{\xi}_{\tilde{n}\tilde{m}\tilde{k}} &= \int b^{h\xi}_{\tilde{n}\tilde{m}\tilde{k}l}\text{d}h_{\tilde{n}\tilde{m}\tilde{k}l}.
	\end{align}
\end{subequations}
Combining \eqref{eq:sec4 FC1} with \eqref{eq:sec4 MaCC1} can yield the MCC about the random variable $\xi_{\tilde{n}\tilde{m}\tilde{k}}$
\begin{align}\label{eq:sec4 MaCC2}
	q^{\xi}_{\tilde{n}\tilde{m}\tilde{k}} = b^{\xi}_{\tilde{n}\tilde{m}\tilde{k}} = g^{\xi}_{\tilde{n}\tilde{m}\tilde{k}}.
\end{align}

In large-scale systems, verifying MCCs concerning continuous random variables is too complicated to derive tractable messages. 
Considering that many variables in practical systems can be characterized or approximated by exponential family distributions, we constrain the beliefs concerning the random variables $g_{nmkl}$ and $h_{\tilde{n}\tilde{m}\tilde{k}l}$ with Gaussian distributions according to the large-scale condition. 
Therefore, the MCCs concerning $g_{nmkl}$ and $h_{\tilde{n}\tilde{m}\tilde{k}l}$ can be relaxed into the following mean and variance consistency constraints (MVCCs) \cite{cespedes2014expectation}
\begin{subequations}\label{eq:sec4 MoCC}
	\begin{align}
		&\text{E}[g_{nmkl}\vert q^{g}_{nmkl}]  =  \text{E}[g_{nmkl}\vert b^{g}_{nmkl}]  =  \text{E}[g_{nmkl}\vert b^{g\mathbf{h}}_{nmkl}],\\
		&\text{V}[g_{nmkl}\vert q^{g}_{nmkl}]  =  \text{V}[g_{nmkl}\vert b^{g}_{nmkl}]  \hspace{-1mm}=  \text{V}[g_{nmkl}\vert b^{g\mathbf{h}}_{nmkl}],\\
		&\text{E}[h_{\tilde{n}\tilde{m}\tilde{k}l}\vert q^{h}_{\tilde{n}\tilde{m}\tilde{k}l}]  =  \text{E}[h_{\tilde{n}\tilde{m}\tilde{k}l}\vert b^{h}_{\tilde{n}\tilde{m}\tilde{k}l}]  =  \text{E}[h_{\tilde{n}\tilde{m}\tilde{k}l}\vert b^{g\mathbf{h}}_{nmkl}],\\
		& \text{V}[h_{\tilde{n}\tilde{m}\tilde{k}l}\vert q^{h}_{\tilde{n}\tilde{m}\tilde{k}l}]  =  \text{V}[h_{\tilde{n}\tilde{m}\tilde{k}l}\vert b^{h}_{\tilde{n}\tilde{m}\tilde{k}l}] \nonumber \\
		 &\textstyle \qquad \qquad \qquad\quad=  \frac{1}{NMK}\sum_{n,m,k}\text{V}[h_{\tilde{n}\tilde{m}\tilde{k}l}\vert b^{g\mathbf{h}}_{nmkl}].
	\end{align}
\end{subequations}
Since the variable $ h_{\tilde{n}\tilde{m}\tilde{k}l} $ is contained in the belief $ b^{g\mathbf{h}}_{nmkl}$ for any index $n,m,k$, we approximate the corresponding variance consistency constraint for tractability while retaining the original mean consistency constraint for fidelity. 

By combining the above constraint design, the JCEP problem can be transformed into the BFE minimization problem with hybrid constraints, which is given by
\begin{align}\label{eq:sec4 Bethe free energy mini}
    &\text{min}\:{{\mathcal{F}}_{\text{B}}}\quad\text{s.t.}\quad\left\lbrace
    \begin{array}{ll}
        \eqref{eq:sec4 FC1}\eqref{eq:sec4 FC2}\eqref{eq:sec4 FC3},&\text{(FCs)},\\
        \eqref{eq:sec4 MaCC1}\eqref{eq:sec4 MaCC2},&\text{(MCCs)},\\
        \eqref{eq:sec4 MoCC},&\text{(MVCCs)}.
    \end{array}
    \right.
\end{align}

We resort to the Lagrange multipliers method to solve the above optimization problem with hybrid constraints. 
Substituting the FCs into \eqref{eq:bethe free energy}, BFE can be rewritten as \eqref{eq:BFE2}. 
The corresponding Lagrange function is formulated as \eqref{eq: lagrange}. 
\begin{figure*}[h]
	\begin{align}\label{eq:BFE2}
		\tilde{\mathcal{F}}_{\text{B}} = &\textstyle\sum_{n,m,k,l}\text{D}[b^{g}_{nmkl}\Vert p(y_{nmkl}\vert g_{nmkl})]
		+ 
		\sum_{n,m,k,l}\text{D}
		[ b^{g\mathbf{h}}_{nmkl} \Vert p(g_{nmkl}\vert\mathbf{h}_{l};\hat{\boldsymbol{\omega}} )]
		+
		\sum_{\tilde{n},\tilde{m},\tilde{k},l}\text{D}[b^{h\xi}_{\tilde{n}\tilde{m}\tilde{k}l} \Vert p( h_{\tilde{n}\tilde{m}\tilde{k}l}\vert\xi_{\tilde{n}\tilde{m}\tilde{k}};\hat{\sigma}_{\tilde{n}\tilde{m}\tilde{k}} )]
		\nonumber\\
		&\textstyle+
		\sum_{\tilde{n},\tilde{m},\tilde{k}}\text{D}[b^{\xi}_{\tilde{n}\tilde{m}\tilde{k}}\Vert p( \xi_{\tilde{n}\tilde{m}\tilde{k}};\hat{\rho} )]
		+
		\sum_{n,m,k,l}\text{H}[q^{g}_{nmkl}]
		+
		\sum_{\tilde{n},\tilde{m},\tilde{k}}\text{H}[q^{\xi}_{\tilde{n}\tilde{m}\tilde{k}}]
		+
		\sum_{\tilde{n},\tilde{m},\tilde{k},l}NMK\text{H}[q^{h}_{\tilde{n}\tilde{m}\tilde{k}l}]
		.
	\end{align}
	\hrulefill
\end{figure*}
\begin{figure*}[!htb]
	\begin{align}\label{eq: lagrange}
		\mathcal{L}_{\text{B}}& = \textstyle\tilde{\mathcal{F}}_{\text{B}} 
		+   
		\sum_{\tilde{n},\tilde{m},\tilde{k},l}2\text{Re}\left\lbrace \left(\alpha^{h,b^{h}}_{\tilde{n}\tilde{m}\tilde{k}l}\right)^{*}\left(
		\text{E}\left[h_{\tilde{n}\tilde{m}\tilde{k}l}\vert q^{h}_{\tilde{n}\tilde{m}\tilde{k}l}\right] - \text{E}\left[h_{\tilde{n}\tilde{m}\tilde{k}l}\vert b^{h}_{\tilde{n}\tilde{m}\tilde{k}l}\right]
		\right) \right\rbrace \nonumber\\
		&\textstyle
		+    
		\sum_{\tilde{n},\tilde{m},\tilde{k},l}\sum_{n,m,k}2\text{Re}\left\lbrace \left(\alpha^{h,b^{g\textbf{h}}}_{nmk,\tilde{n}\tilde{m}\tilde{k}l}\right)^{*}\left(
		\text{E}\left[h_{\tilde{n}\tilde{m}\tilde{k}l}\vert q^{h}_{\tilde{n}\tilde{m}\tilde{k}l}\right] - \text{E}\left[h_{\tilde{n}\tilde{m}\tilde{k}l}\vert b^{g\mathbf{h}}_{nmkl}\right]
		\right) \right\rbrace \nonumber\\
		&\textstyle 
		+
		\sum_{\tilde{n},\tilde{m},\tilde{k},l}\left(\beta^{h,b^{h}}_{\tilde{n}\tilde{m}\tilde{k}l}\left(\text{V}\hspace{-1mm}\left[h_{\tilde{n}\tilde{m}\tilde{k}l}\vert q^{h}_{\tilde{n}\tilde{m}\tilde{k}l}\right]  \hspace{-1mm}-\hspace{-1mm} \text{V}\hspace{-1mm}\left[h_{\tilde{n}\tilde{m}\tilde{k}l}\vert b^{h}_{\tilde{n}\tilde{m}\tilde{k}l}\right]\right) +  
		\beta^{h,b^{g\textbf{h}}}_{\tilde{n}\tilde{m}\tilde{k}l}\left(NMK\text{V}\hspace{-1mm}\left[h_{\tilde{n}\tilde{m}\tilde{k}l}\vert q^{h}_{\tilde{n}\tilde{m}\tilde{k}l}\right] \hspace{-1mm}-\hspace{-1mm} \sum_{n,m,k}\text{V}\hspace{-1mm}\left[h_{\tilde{n}\tilde{m}\tilde{k}l}\vert b^{g\mathbf{h}}_{nmkl}\right]\right)\right) \nonumber\\
		&\textstyle +   
		\sum_{n,m,k,l}2\text{Re}\left\lbrace
		\left(\alpha^{g,b^{g}}_{nmkl}\right)^{*}
		\left(\text{E}\left[g_{nmkl}\vert q^{g}_{nmkl}\right]  -  \text{E}\left[g_{nmkl}\vert b^{g}_{nmkl}\right]\right) + 
		\left(\alpha^{g,b^{g\textbf{h}}}_{nmkl}\right)^{*}
		\left(\text{E}\left[g_{nmkl}\vert q^{g}_{nmkl}\right]  -  \text{E}\left[g_{nmkl}\vert b^{g\mathbf{h}}_{nmkl}\right]\right)
		\right\rbrace
		\nonumber\\
		&\textstyle +   
		\sum_{n,m,k,l}\left(\beta^{g,b^{g}}_{nmkl}
		\left(\text{V}\left[g_{nmkl}\vert q^{g}_{nmkl}\right] - \text{V}\left[g_{nmkl}\vert b^{g}_{nmkl}\right] + \beta^{g,b^{g\textbf{h}}}_{nmkl}\left(\text{V}\left[g_{nmkl}\vert q^{g}_{nmkl}\right] - \text{V}\left[g_{nmkl}\vert b^{g\mathbf{h}}_{nmkl}\right] \right)  \right) \right)
		\nonumber\\
		&\textstyle +   
		\sum_{\tilde{n},\tilde{m},\tilde{k}}\sum_{\xi_{\tilde{n}\tilde{m}\tilde{k}}\in\mathbb{B}}\nu^{\xi}_{\tilde{n}\tilde{m}\tilde{k}}
		\left(b^{\xi}_{\tilde{n}\tilde{m}\tilde{k}} - \int b^{h\xi}_{\tilde{n}\tilde{m}\tilde{k}l}\text{d}h_{\tilde{n}\tilde{m}\tilde{k}l}\right)
		+  
		\sum_{\tilde{n},\tilde{m},\tilde{k},l}\int \nu^{h}_{\tilde{n}\tilde{m}\tilde{k}l}
		\left(b^{h}_{\tilde{n}\tilde{m}\tilde{k}l}
		-\sum_{\xi_{\tilde{n}\tilde{m}\tilde{k}}\in\mathbb{B}}b^{h\xi}_{\tilde{n}\tilde{m}\tilde{k}l}\right)
		\text{d}h_{\tilde{n}\tilde{m}\tilde{k}l}.
	\end{align}
	\hrulefill
\end{figure*}
Setting the first-order derivatives of the Lagrange function with respect to the auxiliary beliefs and simplifying, we can derive the Off-Grid-MS HMP algorithm summarized in Algorithm \ref{alg:Off-Grid-MS HMP}, where 
$w_{nmk}^{\tilde{n}\tilde{m}\tilde{k}}\triangleq[\mathbf{W}(\hat{\boldsymbol{\omega}})]_{nMK+mK+k,\tilde{n}\tilde{M}\tilde{K}+\tilde{m}\tilde{K}+\tilde{k}}$.
The derivation of hyper-parameter learning is detailed given in Section \ref{sec low-complexity hyper-parameter}.

\begin{algorithm}[!t]
	\small
	\SetAlgoNoLine
    \caption{Off-Grid-MS HMP Algorithm}
    \label{alg:Off-Grid-MS HMP}
        \textbf{Input:} $\mathbf{Y}$, $\mathbf{S}$, ${\sigma}_{\text{z}}$, $T_{\text{out}}$, $T_{\text{in}}$, $\kappa$. \\
        \textbf{Output:} $\text{LLR}_{\tilde{n}\tilde{m}\tilde{k}}$, $\text{E}[h_{\tilde{n}\tilde{m}\tilde{k}l}\vert b^{h}_{\tilde{n}\tilde{m}\tilde{k}l}]$.\\
        \textbf{Initialize:} $\mu^{h}_{\tilde{n}\tilde{m}\tilde{k}l}=0$, $\tau^{h}_{\tilde{n}\tilde{m}\tilde{k}l}=1$, $\alpha^{g,b^{g}}_{nmkl}=0$, $\beta^{h,b^{h}}_{\tilde{n}\tilde{m}\tilde{k}l}=1$, $\zeta_{\tilde{n}\tilde{m}\tilde{k}}=0.5$, $\hat{\boldsymbol{\omega}} = \mathbf{0}_{\tilde{N}\tilde{M}\tilde{K}}$, 
		$ \hat{\boldsymbol{\sigma}}=\mathbf{1}_{\tilde{N}\tilde{M}\tilde{K}} $, $ \hat{\rho} = 0.2 $.\\
        \For{$t_{\rm{out}}=1,\cdots T_{\rm{out}}$}{
			\For{$t_{\rm{in}}=1,\cdots T_{\rm{in}}$}{
			$\forall n,m,k,l:$\\
			$ \tau^{q}_{nmkl}=\sum_{\tilde{n},\tilde{m},\tilde{k}}\vert w_{nmk}^{\tilde{n}\tilde{m}\tilde{k}}\vert^{2}/\beta^{h,b^{h}}_{\tilde{n}\tilde{m}\tilde{k}l}$;\\
			$\mu^{q}_{nmkl}=-\alpha^{g,b^{g}}_{nmkl}\tau^{q}_{nmkl}+ \sum_{\tilde{n},\tilde{m},\tilde{k}}w_{nmk}^{\tilde{n}\tilde{m}\tilde{k}}\mu^{h}_{\tilde{n}\tilde{m}\tilde{k}l}$;\\
			$b^{g}_{nmkl}\propto p(y_{nmkl}\vert g_{nmkl})\mathcal{CN}(g_{nmkl};\mu^{q}_{nmkl},\tau^{q}_{nmkl})$;\\
			$\mu^{g}_{nmkl}\hspace{-1mm}=\hspace{-1mm}\text{E}\left[g_{nmkl}\vert b^{g}_{nmkl}\right],\tau^{g}_{nmkl}\hspace{-1mm}=\hspace{-1mm}\text{V}\left[g_{nmkl}\vert b^{g}_{nmkl}\right]$;\\
			$\varepsilon_{nmkl}=1/\tau^{q}_{nmkl}-\tau^{g}_{nmkl}/{(\tau^{q}_{nmkl})^{2}}$;\\
			$\alpha^{g,b^{g}}_{nmkl}=(\mu^{g}_{nmkl}-\mu^{q}_{nmkl})/\tau^{q}_{nmkl}$.\\
			$\forall \tilde{n},\tilde{m},\tilde{k},l:$\\
			$\tau^{r}_{\tilde{n}\tilde{m}\tilde{k}l}=(\sum_{n,m,k}\vert w_{nmk}^{\tilde{n}\tilde{m}\tilde{k}}\vert^{2}\varepsilon_{nmkl})^{-1}-(NMK\beta^{h,b^{h}}_{\tilde{n}\tilde{m}\tilde{k}l})^{-1}$;\\
			$\beta^{h,b^{h}}_{\tilde{n}\tilde{m}\tilde{k}l}=(\tau^{h}_{\tilde{n}\tilde{m}\tilde{k}l})^{-1}-(NMK{\tau}_{\tilde{n}\tilde{m}\tilde{k}l}^{r})^{-1}$;\\
			$\mu^{r}_{\tilde{n}\tilde{m}\tilde{k}l}=\mu^{h}_{\tilde{n}\tilde{m}\tilde{k}l}+\tau^{r}_{\tilde{n}\tilde{m}\tilde{k}l}\sum_{n,m,k}(w_{nmk}^{\tilde{n}\tilde{m}\tilde{k}})^{*}\alpha^{g,b^{g}}_{nmkl}$;\\
			$p^{\prime}(h_{\tilde{n}\tilde{m}\tilde{k}l})\propto \mathcal{CN}( h_{\tilde{n}\tilde{m}\tilde{k}l};0,\hat{\sigma}_{\tilde{n}\tilde{m}\tilde{k}})\mathcal{CN}(h_{\tilde{n}\tilde{m}\tilde{k}l};\mu^{r}_{\tilde{n}\tilde{m}\tilde{k}l},\tau^{r}_{\tilde{n}\tilde{m}\tilde{k}l}) $;\\
			$b^{h}_{\tilde{n}\tilde{m}\tilde{k}l}=\zeta_{\tilde{n}\tilde{m}\tilde{k}}p^{\prime}(h_{\tilde{n}\tilde{m}\tilde{k}l})+(1-\zeta_{\tilde{n}\tilde{m}\tilde{k}})\delta(h_{\tilde{n}\tilde{m}\tilde{k}l})$;\\
			$\mu^{h}_{\tilde{n}\tilde{m}\tilde{k}l}=\text{E}[h_{\tilde{n}\tilde{m}\tilde{k}l}\vert b^{h}_{\tilde{n}\tilde{m}\tilde{k}l}],\tau^{h}_{\tilde{n}\tilde{m}\tilde{k}l}=\text{V}[h_{\tilde{n}\tilde{m}\tilde{k}l}\vert b^{h}_{\tilde{n}\tilde{m}\tilde{k}l}]$.	\\
			$\forall \tilde{n},\tilde{m},\tilde{k}:$\\
			$\text{LLR}_{\tilde{n}\tilde{m}\tilde{k}}\hspace{-1mm}=\hspace{-1mm}\ln\frac{\hat{\rho}}{1-\hat{\rho}}\hspace{-1mm}+\hspace{-1mm}\sum_{l}\ln\frac{\mathcal{CN}(\mu^{r}_{\tilde{n}\tilde{m}\tilde{k}l};0,\tau^{r}_{\tilde{n}\tilde{m}\tilde{k}l}+\hat{\sigma}_{\tilde{n}\tilde{m}\tilde{k}})}{\mathcal{CN}(\mu^{r}_{\tilde{n}\tilde{m}\tilde{k}l};0,\tau^{r}_{\tilde{n}\tilde{m}\tilde{k}l})}$;\\
			$\zeta_{\tilde{n}\tilde{m}\tilde{k}}=1-\frac{1}{1+\text{exp}\lbrace{\text{LLR}_{\tilde{n}\tilde{m}\tilde{k}}}\rbrace}$.\\
			update $ \hat{\rho} $ and $\hat{\boldsymbol{\sigma}} $;\\
			}
			update $ \hat{\boldsymbol{\omega}}$;\\
		}
\end{algorithm}

The resulting channel estimator can be represented as
\begin{align}
	\hat{h}_{\tilde{n}\tilde{m}\tilde{k}l} = \text{G}(\text{LLR}_{\tilde{n}\tilde{m}\tilde{k}}-\text{LLR}_{\text{thr}})\text{E}[h_{\tilde{n}\tilde{m}\tilde{k}l}\vert b^{h}_{\tilde{n}\tilde{m}\tilde{k}l}],
\end{align}
where $\text{G}(\cdot)$ denotes the Heaviside step function and $\text{LLR}_{\text{thr}}$ is the predefined log-likelihood ratio (LLR). 

\textit{Remark 4:} In the large-scale systems, $(NMK\beta^{h,b^{h}}_{\tilde{n}\tilde{m}\tilde{k}l})^{-1}$ and $(NMK{\tau}_{\tilde{n}\tilde{m}\tilde{k}l}^{r})^{-1}$ in Steps 14 and 15 of Algorithm \ref{alg:Off-Grid-MS HMP} approach to zero. Therefore, the Off-Grid-MS HMP algorithm without the DAD-domain off-grid model degenerates into the EM-BG-AMP-MMV algorithm \cite{chen2018sparse} or EM-BG-AMP algorithm \cite{vila2013expectation}, depending on whether the common channel statistical characteristics of different subbands are considered or not.

\textit{Remark 5:} Excluding the complexity of hyper-parameter learning which will be discussed in Section \ref{sec low-complexity hyper-parameter}, the complexity of the HMP algorithm is $\mathcal{O}(NMK\tilde{N}\tilde{M}\tilde{K}L)$, primarily dominated by the complicated matrix multiplication. Motivated by the sparsity of the DAD-domain channel, the HMP algorithm can be operated on the active DAD-domain grids whose LLRs of state detectors are greater than the predefined LLR threshold. The numbers of active grids in the delay, angle, and Doppler domains are denoted by $\bar{N}$, $\bar{M}$, and $\bar{K}$, respectively. Note that the number of total grids is much larger than that of active grids, i.e., $\tilde{N}\tilde{M}\tilde{K}\gg \bar{N}\bar{M}\bar{K}$. After pruning, the complexity of the HMP algorithm can be reduced to $\mathcal{O}(NMK\bar{N}\bar{M}\bar{K}L)$.

\subsection{Low-Complexity Hyper-Parameter Learning}
\label{sec low-complexity hyper-parameter}
In practice, hyper-parameters vary depending on the channel conditions and are typically unknown to the BS. The proposed Off-Grid-MS HMP algorithm can adaptively learn the hyper-parameters of the channel. Preserving the terms of the Lagrange function \eqref{eq: lagrange} related to the hyper-parameters, we can obtain the corresponding optimization problems. 

Firstly, we consider the learning of $ \rho $ whose optimization problem can be denoted as 
\begin{align}
	\label{eq:EM rho}
	\textstyle
	\hat{\rho} = \underset{\rho}{\text{arg max}}\:
	\text{E}\left[
	\sum_{\tilde{n},\tilde{m},\tilde{k}}b^{\xi}_{\tilde{n}\tilde{m}\tilde{k}}\ln p( \xi_{\tilde{n}\tilde{m}\tilde{k}};\rho )
	\right].
\end{align}
By setting the first-order derivative of \eqref{eq:EM rho} concerning $ \rho $ to zero, we can obtain the estimation of $ \rho $ as 
\begin{align}
	\textstyle
	\hat{\rho} = \frac{1}{\tilde{N}\tilde{M}\tilde{K}}\sum_{\tilde{n},\tilde{m},\tilde{k}}\zeta_{\tilde{n}\tilde{m}\tilde{k}}.
\end{align}
Similarly, the optimization problem concerning the hyper-parameter $ \sigma_{\tilde{n}\tilde{m}\tilde{k}} $ is given by
\begin{align}
	\label{eq:EM sigma}
	\textstyle
	\hat{\sigma}_{\tilde{n}\tilde{m}\tilde{k}} \hspace{-1mm}=\hspace{-1mm} \underset{\sigma_{\tilde{n}\tilde{m}\tilde{k}}}{\text{arg max}}\:\text{E}\hspace{-0.5mm}\left[
	\sum_{l}\hspace{-0.5mm}
	b^{h}_{{\tilde{n}\tilde{m}\tilde{k}l}}
	\ln p( h_{\tilde{n}\tilde{m}\tilde{k}l}\vert\xi_{\tilde{n}\tilde{m}\tilde{k}};\sigma_{\tilde{n}\tilde{m}\tilde{k}} ) \hspace{-0.5mm}\right].
\end{align}
By taking the first-order derivative of \eqref{eq:EM sigma} concerning $ \sigma_{\tilde{n}\tilde{m}\tilde{k}} $ to zero, we can obtain
\begin{align}
	\textstyle
	\hat{\sigma}_{\tilde{n}\tilde{m}\tilde{k}} = \frac{1}{L}\sum_{l}\left(
	\vert\mu^{h}_{\tilde{n}\tilde{m}\tilde{k}l}\vert^{2} + 
	\tau^{h}_{\tilde{n}\tilde{m}\tilde{k}l}\right).
\end{align}

In terms of the off-grid hyper-parameters, the corresponding optimization problem can be formulated as
\begin{align}
	\hat{\mathbf{x}} &= \underset{\mathbf{x}}{\text{arg max}}\:\text{E}\Bigg[
	\sum_{l}\Big(\hspace{-1mm}
	\prod_{n,m,k}b^{g\mathbf{h}}_{nmkl}\Big)\ln \Big(\hspace{-1mm}\prod_{n,m,k}p(g_{nmkl}\vert\mathbf{h}_{l};\boldsymbol{\omega} )\Big)
	\Bigg] \nonumber\\
	\label{eq:EM off-grid2}
	&=\underset{\mathbf{x}}{\text{arg max}}\:
	\mathbf{x}^{T}\boldsymbol{\Xi}_{x}\mathbf{x} - 2\boldsymbol{\chi}_{x}^{T}\mathbf{x},
\end{align}
where $\mathbf{x}\in\lbrace \boldsymbol{\alpha},\boldsymbol{\beta}, \boldsymbol{\gamma},\boldsymbol{\eta}\rbrace$ and $x\in\lbrace \alpha,\beta,\gamma,\eta \rbrace$.
The matrix $ \boldsymbol{\Xi}_{x} $ and the vector $ \boldsymbol{\chi}_{x} $ are denoted as
\begin{subequations}
	\label{eq:PiChi alpha}
	\begin{align}
		\label{eq:Pi alpha}
		\boldsymbol{\Xi}_{x} =& \mathbf{R}_{x}^{T}\text{Re}\Big\lbrace
		\Big(
		\dot{\mathbf{W}}_{x}^{H}
		\dot{\mathbf{W}}_{x}
		\Big)^{*}
		\odot
		\mathbf{U}
		\Big\rbrace\mathbf{R}_{x},\\
		\label{eq:Chi alpha}
		\boldsymbol{\chi}_{x} =& \mathbf{R}_{x}^{T}\text{Re}\Big\lbrace
		\frac{1}{L}( (\mathbf{M}^{h})^{*} \odot
		\mathbf{V}^{x})\mathbf{1}_{L}
		\Big\rbrace \nonumber \\
		& -
		\mathbf{R}_{x}^{T}\text{Re}\Big\lbrace
		\text{diag}\Big\lbrace
		(
		\dot{\mathbf{W}}_{x}^{H}
		\mathbf{W}_{\backslash x}
		)
		\boldsymbol{\Sigma}^{h}
		\Big\rbrace
		\Big\rbrace.
	\end{align}
\end{subequations}
The posterior mean matrix of $\mathbf{H}$ is $ \mathbf{M}^{h}=[ \boldsymbol{\mu}_{0}^{h},\boldsymbol{\mu}_{1}^{h},\cdots,\boldsymbol{\mu}_{L-1}^{h} ] $ whose the $(\tilde{n}\tilde{M}\tilde{K} + \tilde{m}\tilde{K} + \tilde{k},l)$-th element is $\mu_{\tilde{n}\tilde{m}\tilde{k}l}^{h}$. The posterior covariance matrix of $\mathbf{H}$ is $ \boldsymbol{\Sigma}^{h} $. It is a diagonal matrix whose the $ (\tilde{n}\tilde{M}\tilde{K} + \tilde{m}\tilde{K} + \tilde{k}) $-th diagonal element is $ \frac{1}{L}\sum_{l}\tau_{\tilde{n}\tilde{m}\tilde{k}l}^{h} $. For the expression simplicity, we define 
$\mathbf{W}_{\backslash x}\triangleq \mathbf{W}(\boldsymbol{\omega}) - \dot{\mathbf{W}}_{x} $.
The matrix $ \mathbf{U}\triangleq\frac{1}{L}\mathbf{M}^{h}(\mathbf{M}^{h})^{H} + \boldsymbol{\Sigma}^{h}$. 
The matrix $ \mathbf{V}^{x}\triangleq\dot{\mathbf{W}}_{x}^{H}( \mathbf{M}^{g}-\mathbf{W}_{\backslash x}\mathbf{M}^{h} ) $ whose $ (\tilde{n}\tilde{M}\tilde{K}+\tilde{n}\tilde{k}+\tilde{k},l) $-th element is $ v^{x}_{\tilde{n}\tilde{m}\tilde{k}l} $. 
The posterior mean matrix of $\mathbf{G}$ is $ \mathbf{M}^{g}=[ \boldsymbol{\mu}_{0}^{g},\boldsymbol{\mu}_{1}^{g},\cdots,\boldsymbol{\mu}_{L-1}^{g} ] $ whose the $(nMK + mK + k,l)$-th element is $\mu_{nmkl}^{g}$. 
The detailed derivation of \eqref{eq:EM off-grid2} is given in Appendix \ref{Appendix A}. By setting the first-order derivative of \eqref{eq:EM off-grid2} with respect to $ \mathbf{x} $ to zero, we can obtain the estimation of off-grid hyper-parameter vector, denoted as $ \hat{\mathbf{x}} = \boldsymbol{\Xi}_{x}^{-1}\boldsymbol{\chi}_{x} $.

Based on the above expressions of hyper-parameter estimation, we first analyze their computational complexity, as shown in Table \ref{tab:complexity hyper}. It can be observed that off-grid hyper-parameter learning introduces significantly higher computational complexity than others, primarily dominated by matrix multiplication. Therefore, we focus attention on reducing the complexity of off-grid hyper-parameter learning.
\begin{table}[!ht]
	\small
	\centering
	\caption{Computational complexity of hyper-parameter learning.}
	\label{tab:complexity hyper}
	\begin{tabular}{cc}
		\hline
		\textbf{Hyper-parameter} & \textbf{ Complexity}\\ \hline
		$ \rho $ 
		& $ \mathcal{O}(\tilde{N}\tilde{M}\tilde{K}) $   \\ 
		$ \boldsymbol{\sigma} $ 
		& $ \mathcal{O}(\tilde{N}\tilde{M}\tilde{K}L) $   \\ 
		Original $ \boldsymbol{\alpha},\boldsymbol{\beta},\boldsymbol{\gamma},\boldsymbol{\eta} $
		& $ \mathcal{O}(\tilde{N}^{2}\tilde{M}^{2}\tilde{K}^{2}L) $   \\ 
		Simplified $ \boldsymbol{\alpha} $
		& 
		$\mathcal{O}(\tilde{K}^{2}\tilde{N}^{2}\tilde{M}L) $   
		\\ 
		Simplified $ \boldsymbol{\beta} $
		& 
		$\mathcal{O}(\tilde{K}^{2}\tilde{M}_{\text{v}}\tilde{N}\tilde{M}L) $
		\\ 
		Simplified $ \boldsymbol{\gamma} $
		& 
		$\mathcal{O}(\tilde{K}^{2}\tilde{M}_{\text{h}}\tilde{N}\tilde{M}L) $
		\\ 
		Simplified $ \boldsymbol{\eta} $
		& 
		$\mathcal{O}( \tilde{K}^{2}\tilde{N}\tilde{M}L ) $  
		\\ \hline
	\end{tabular}
\end{table}

As mentioned in Section \ref{subsec:off-grid basis}, we set $ \tilde{K}>K $ to enhance time-domain channel prediction, while setting $ \tilde{N}=N $, $ \tilde{M}_{\text{v}}=M_{\text{v}} $, and $ \tilde{M}_{\text{h}}=M_{\text{h}} $ for the approximations $ \mathbf{B}^{H}(\boldsymbol{\alpha})\mathbf{B}(\boldsymbol{\alpha})\approx N\mathbf{I}_{\tilde{N}} $, $ \mathbf{C}_{\text{v}}^{H}(\boldsymbol{\beta})\mathbf{C}_{\text{v}}(\boldsymbol{\beta})\approx M_{\text{v}}\mathbf{I}_{\tilde{M}_{\text{v}}} $, and $ \mathbf{C}_{\text{h}}^{H}(\boldsymbol{\gamma})\mathbf{C}_{\text{h}}(\boldsymbol{\gamma})\approx M_{\text{h}}\mathbf{I}_{\tilde{M}_{\text{h}}} $. 
Building upon the above approximations, we can simplify the computation of off-grid hyper-parameter learning. 

Starting from the hyper-parameter $ \boldsymbol{\eta} $, 
the matrix $ \boldsymbol{\Xi}_{\eta} $ can be approximated as
\begin{align}
	\boldsymbol{\Xi}_{\eta}
	&\overset{(\text{a})}{\approx}\mathbf{R}_{\eta}^{T}
	\text{Re}\big\lbrace
	( NM\mathbf{I}_{\tilde{N}\tilde{M}} \otimes ( \dot{\mathbf{D}}^{H}\dot{\mathbf{D}} ) )^{*}
	\odot \mathbf{U}
	\big\rbrace\mathbf{R}_{\eta} \nonumber\\
	&\overset{(\text{b})}{=}NM
	\text{Re}\Big\lbrace
	( \dot{\mathbf{D}}^{H}\dot{\mathbf{D}} )^{*} \odot
	\mathbf{U}^{\eta}
	\Big\rbrace,
\end{align}
where approximation (a) is from the approximations of the off-grid matrices.
Equation (b) is based on the definitions of the matrix $ \mathbf{R}_{\eta} $ and the block diagonal matrix $ \mathbf{I}_{\tilde{N}M} \otimes ( \dot{\mathbf{D}}^{H}\dot{\mathbf{D}} ) $. 
The matrix $\dot{\mathbf{D}}$ is the first-order derivation of $ \mathbf{D}\triangleq\mathbf{D}(\mathbf{0}_{\tilde{K}}) $ concerning Doppler shift.
$ \mathbf{U}^{\eta}\triangleq\frac{1}{L}\mathbf{M}^{h}_{\eta}\left( \mathbf{M}^{h}_{\eta} \right)^{H} + \boldsymbol{\Sigma}^{h}_{\eta} $, where the $ (\tilde{k},\tilde{n}\tilde{M}L+\tilde{m}L+l) $-th element of $ \mathbf{M}^{h}_{\eta}\in\mathbb{C}^{\tilde{K}\times \tilde{N}\tilde{M}L} $ is $ {\mu}^{h}_{\tilde{n}\tilde{m}\tilde{k}l} $ and the $ \tilde{k} $-th diagonal element of the diagonal matrix $ \boldsymbol{\Sigma}^{h}_{\eta}\in\mathbb{R}^{\tilde{K}\times\tilde{K}} $ is $ \frac{1}{L}\sum_{\tilde{n},\tilde{m},l}\tau_{\tilde{n}\tilde{m}\tilde{k}l}^{h} $. 
Similarly, the vector $ \mathbf{\boldsymbol{\chi}}_{\eta} $ can be approximated as
\begin{align}
	\mathbf{\boldsymbol{\chi}}_{\eta}
	\approx
	&\text{Re}\Big\lbrace
	\Big(
	\frac{1}{L}( \mathbf{M}^{h}_{\eta} )^{*} \odot \tilde{\mathbf{V}}^{\eta}
	\Big) \mathbf{1}_{\tilde{N}\tilde{M}L}
	\Big\rbrace \nonumber \\ 
	&-NM\text{Re}\big\lbrace
	\text{diag}\lbrace (\dot{\mathbf{D}}^{H}\mathbf{D}) \odot \boldsymbol{\Sigma}^{h}_{\eta} \rbrace
	\Big\rbrace,
\end{align}
where the $ (\tilde{k},\tilde{n}\tilde{M}L+\tilde{m}L+l) $-th element of the matrix $ \tilde{\mathbf{V}}^{\eta}\in\mathbb{C}^{\tilde{K}\times \tilde{N}\tilde{M}L} $ is $ v^{\eta}_{\tilde{n}\tilde{m}\tilde{k}l} $.

Then, we simplify the learning of the hyper-parameter $ \boldsymbol{\alpha} $. the matrix $ \boldsymbol{\Xi}_{\alpha} $ can be approximated as 
\begin{align}
	\boldsymbol{\Xi}_{\alpha} &\hspace{-1mm}\overset{(a)}{\approx}\hspace{-1mm}
	\mathbf{R}_{\alpha}^{T}\text{Re}\Big\lbrace
	\Big(
	(\dot{\mathbf{B}}^{H}\dot{\mathbf{B}}) \hspace{-1mm}\otimes\hspace{-1mm} (M\mathbf{I}_{\tilde{M}}) \hspace{-1mm}\otimes\hspace{-1mm} (\mathbf{D}^{H}(\boldsymbol{\eta})\mathbf{D}(\boldsymbol{\eta}))
	\Big)^{*}
	\hspace{-1mm}\odot\hspace{-1mm}
	\mathbf{U}
	\Big\rbrace\mathbf{R}_{\alpha} \nonumber\\
	&\hspace{-1mm}\overset{(b)}{=}\hspace{-1mm}
	M\tilde{\mathbf{R}}_{\alpha}^{T}
	\text{Re}\Big\lbrace
	\big(
	( \dot{\mathbf{B}}^{H}\dot{\mathbf{B}} ) \hspace{-1mm}\otimes\hspace{-1mm}
	( \mathbf{D}^{H}(\boldsymbol{\eta})\mathbf{D}(\boldsymbol{\eta}) )
	\big)^{*} 
	\hspace{-1mm}\odot\hspace{-1mm} \mathbf{U}^{\alpha}
	\Big\rbrace
	\tilde{\mathbf{R}}_{\alpha},
\end{align}
where approximation (a) is from $ \mathbf{C}^{H}(\boldsymbol{\beta},\boldsymbol{\gamma})\mathbf{C}(\boldsymbol{\beta},\boldsymbol{\gamma})\approx M\mathbf{I}_{\tilde{M}} $, equation (b) is from the definitions of the matrix $\mathbf{R}_{\alpha}$ and the block diagonal matrix $ \mathbf{I}_{\tilde{M}}\otimes \left( \mathbf{D}^{H}(\boldsymbol{\eta})\mathbf{D}(\boldsymbol{\eta}) \right) $. 
The matrix $ \tilde{\mathbf{R}}_{\alpha}\triangleq \mathbf{I}_{\tilde{N}}\otimes \mathbf{1}_{\tilde{K}} $. 
The matrix $\dot{\mathbf{B}}$ is the first-order derivation of $ \mathbf{B}\triangleq\mathbf{B}(\mathbf{0}_{\tilde{N}}) $ concerning delay.
$\mathbf{U}^{\alpha}\triangleq \frac{1}{L}\tilde{\mathbf{M}}_{\alpha}^{h}\big(\tilde{\mathbf{M}}_{\alpha}^{h}\big)^{H}+\tilde{\boldsymbol{\Sigma}}_{\alpha}^{h}$, where the $ (\tilde{n}\tilde{K}+\tilde{k},\tilde{m}L+l) $-th element of $\tilde{\mathbf{M}}_{\alpha}^{h}\in\mathbb{C}^{\tilde{N}\tilde{K}\times \tilde{M}L}$ is $ {\mu}_{\tilde{n}\tilde{m}\tilde{k}l}^{h} $ and the $(\tilde{n}\tilde{K}+\tilde{k})$-th diagonal element of the diagonal matrix $ \tilde{\boldsymbol{\Sigma}}_{\alpha}^{h}\in\mathbb{R}^{\tilde{N}\tilde{K}\times \tilde{N}\tilde{K}} $ is $ \frac{1}{L}\sum_{\tilde{m},l}\tau_{\tilde{n}\tilde{m}\tilde{k}l}^{h} $.
Similarly, the vector $ \boldsymbol{\chi}_{\alpha} $ can be approximated as 
\begin{align}
	\boldsymbol{\chi}_{\alpha}\approx&
	\text{Re}\Big\lbrace
	\Big(
	\frac{1}{L}( \mathbf{M}^{h}_{\alpha} )^{*} \odot \tilde{\mathbf{V}}^{\alpha}
	\Big)\mathbf{1}_{\tilde{M}\tilde{K}L}
	\Big\rbrace  \\
	&- 
	M\tilde{\mathbf{R}}_{\alpha}^{T}\text{Re}\Big\lbrace
	\text{diag}\Big\lbrace
	\big( \dot{\mathbf{B}}^{H}\dot{\mathbf{B}} \big) \hspace{-1mm}\otimes\hspace{-1mm}
	\big( \mathbf{D}^{H}(\boldsymbol{\eta})\mathbf{D}(\boldsymbol{\eta}) \big) \hspace{-1mm}\odot\hspace{-1mm}
	\tilde{\boldsymbol{\Sigma}}^{h}_{\alpha}
	\Big\rbrace
	\Big\rbrace,\nonumber
\end{align}
where the $ (\tilde{n},\tilde{m}\tilde{K}L+\tilde{k}L+l) $-th element of $ \mathbf{M}^{h}_{\alpha}\in\mathbb{C}^{\tilde{N}\times \tilde{M}\tilde{K}L} $ and $ \tilde{\mathbf{V}}^{\alpha}\in\mathbb{C}^{\tilde{N}\times \tilde{M}\tilde{K}L} $ are $ {\mu}^{h}_{\tilde{n}\tilde{m}\tilde{k}l} $ and $ v^{\alpha}_{\tilde{n}\tilde{m}\tilde{k}l} $, respectively. 
The approximate operations for the hyper-parameters $ \boldsymbol{\beta} $ and $ \boldsymbol{\gamma} $ are similar to that of $ \boldsymbol{\alpha} $, and we omit the detailed derivation due to space limitations. Based on the above simplification, the computational complexity of the off-grid hyper-parameters is notably reduced, as listed in Table \ref{tab:complexity hyper}.

\section{Simulation Results}
\label{sec5}
\subsection{Simulation Setup}
In this section, we present the simulation results to illustrate the performance of our proposed algorithm.  
Since QuaDRiGa can generate time-varying massive MIMO-OFDM channels \cite{jaeckel2014quadriga} that meet 3GPP new radio (NR) standards \cite{38.901} and have been validated by various field tests, we adopt QuaDRiGa for channel generation in the 3GPP 3D urban marco (UMa) non-line-of-sight (NLOS) scenario for simulations.
Unless otherwise specified, the basic system parameters of simulations are generated according to 3GPP standards, as shown in Table \ref{tab:system parameters}. 
In addition, the SNR is defined as $\text{SNR} = 10\text{log}_{10}\left(\frac{\Vert \mathbf{G} \Vert_{\text{F}}^{2}}{NMKL\sigma_{\text{z}}}\right)$.
\begin{table}[!tb]
	\small
	\centering
	\caption{Basic System Parameters}
	\label{tab:system parameters}
	\begin{tabular}{cc}
		\hline
		\textbf{System Parameters} & \textbf{Value}  \\ \hline  
		Centering frequency $ f_{\text{c}} $ & 3.5 GHz \\ 
		Bandwidth $ B $  & 100 MHz  \\ 
		FFT size $ N_{\text{FFT}} $ & 4096 \\ 
		Number of subcarriers for transmission  $ N_{\text{SC}} $ & 3264 \\ 
		Subcarrier spacing $ \Delta \phi $ & 30 kHz \\ 
		Number of subbands $ L $ & 4 \\ 
		Number of transmission combs $ N_{\text{TC}} $ & 4 \\ 
		Length of SRS sequences $ N $ & 204  \\ 
		Number of BS antennas $ [M_{\text{v}},M_{\text{h}}] $ & [4,8]  \\ 
		Number of fullband sounding $ K $ & 10  \\ 
		Doppler-domain oversampling factor $ S_{\nu} $ & 3 \\ 
		Time interval between SRSs $ \Delta t $ & 1 symbol  \\ 
		Time interval between fullband soundings $ \Delta T $ & 4 slots  \\ 
		Velocity of UE $ v $ & 60 km/h  \\ \hline
	\end{tabular}
\end{table} 

\subsection{Baselines and Performance Metric} 
To demonstrate the superiority of the proposed algorithm, we compare it against the following baselines:
\begin{itemize}
	\item \textbf{DN-LS:} This algorithm estimates the FSTCRM by the LS algorithm and denoises the estimated channel in the delay-angle-time domain with a predefined threshold.
	\item \textbf{HHMP}\cite{liu2021sparse}: This algorithm estimates the delay-angle-time-domain channel by incorporating a hidden Markov model to capture the structured sparsity and temporal dependency characteristic. 
	\item \textbf{OMP}\cite{tropp2007signal}: This algorithm estimates the DADCRM under the SS signal model \eqref{eq:received signal model} without the off-grid model.
	\item \textbf{SOMP}\cite{determe2015exact}: This algorithm is the MMV version of OMP under the MS signal model \eqref{eq:received signal model2}, considering the common channel statistical characteristics of different subbands;
	\item \textbf{EM-BG-AMP}\cite{vila2013expectation}: This algorithm assumes an unknown BG prior, learns the hyper-parameters through the EM, and estimates the DADCRM by the AMP under the SS signal model \eqref{eq:received signal model} without the off-grid model;
	\item \textbf{EM-BG-AMP-MMV}\cite{chen2018sparse}: This algorithm is the MMV version of EM-BG-AMP under the MS signal model \eqref{eq:received signal model2}, considering the common channel statistical characteristics of different subbands.
\end{itemize}
Since DN-LS and HHMP estimate the time-domain channels, we adopt the AR algorithm \cite{lv2019channel} and the PDA algorithm \cite{yin2020addressing} based on the delay-angle-time-domain channel estimation to predict future channels. 
Note that the aforementioned channel prediction baselines can only predict the channels on the pilot symbols. 
Hence, we employ the MMSE interpolation method to predict the channels on the non-pilot symbols, assuming that the BS possesses prior knowledge of maximum Doppler shift and SNR \cite{dong2007linear}. 
In addition, since other baselines can estimate the Doppler-domain channels, they can predict the channels through extrapolation according to the channel model.

To illustrate the performance of channel estimation and prediction, we define the NMSE performance metric 
\begin{align}
	\textstyle\text{NMSE} =
	10\text{log}_{10} \left(\frac{\Vert\hat{\mathbf{G}}-\mathbf{G}\Vert_{\text{F}}^{2}}{\Vert\mathbf{G}\Vert_{\text{F}}^{2}}\right),
\end{align}
where $ \mathbf{G} $ and $ \hat{\mathbf{G}} $ represent the actual and estimated or predicted FSTCRM. 
Note that the predicted FSTCRM in the time domain includes the predicted channels from the last pilot symbol in the current to the first pilot symbol in the future.  

\begin{figure}[!tb]
	\centering
	\subfigure[Channel estimation]{
		\includegraphics[width=0.85\linewidth]{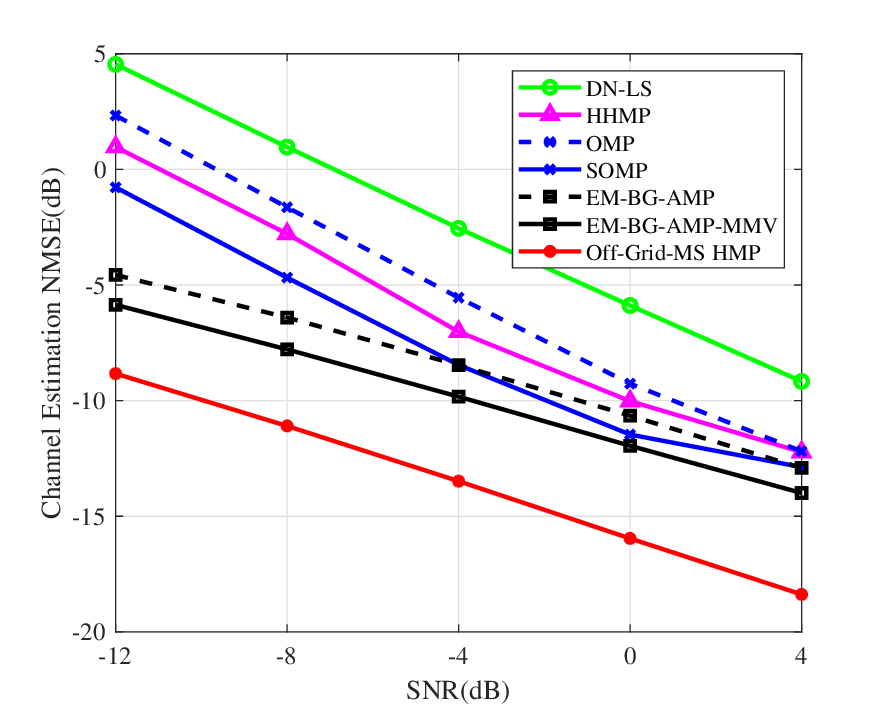}
	}
	\subfigure[Channel prediction]{
		\includegraphics[width=0.85\linewidth]{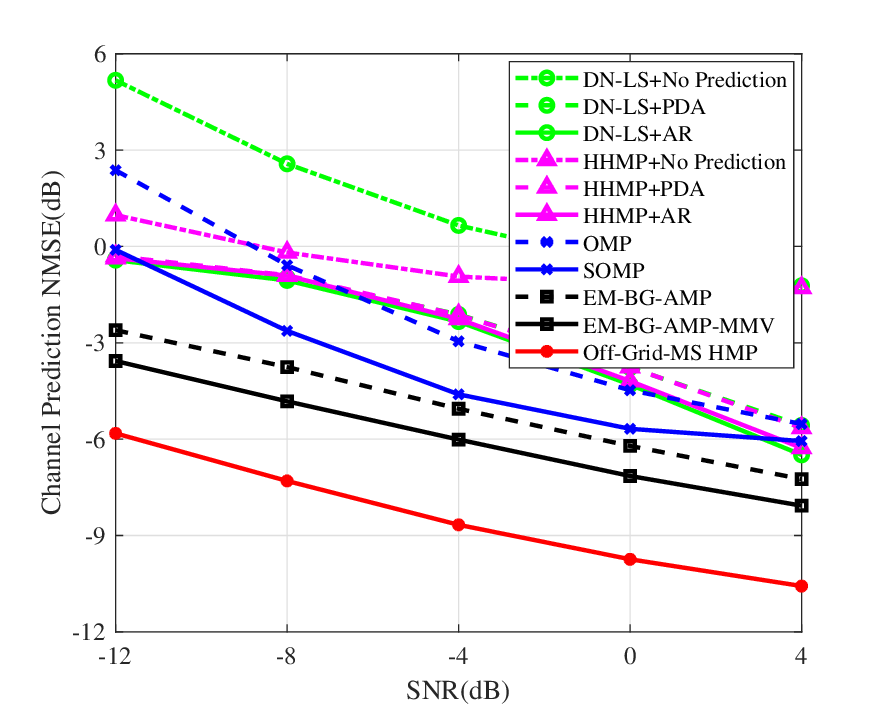}
	}	
	\caption{The performance of JCEP versus SNR.}
	\label{fig:simu SNR}
\end{figure}

\subsection{Simulation Results}
The NMSE of the proposed algorithm and the baselines with the variation of SNR is shown in Fig. \ref{fig:simu SNR}. 
It can be seen that the proposed Off-Grid-MS HMP algorithm significantly outperforms baselines at the entire SNR region by addressing the energy leakage problem through off-grid hyper-parameter learning and utilizing common statistical characteristics across different subbands. 
Benefiting from the learning of channel prior information, EM-BG-AMP/EM-BG-AMP-MMV can achieve better performance than OMP/SOMP.
From the performance comparison of SMV algorithms (OMP and EM-BG-AMP) and MMV algorithms (SOMP and EM-BG-AMP-MMV), it is clear that the common channel statistical characteristics of different subbands can improve the accuracy of CSI acquisition. 
Although HHMP achieves a channel estimation gain of close to 4 dB at the entire SNR region compared to DN-LS, the channel prediction performance of both non-JCEP algorithms is almost identical because the CSI loss between channel estimation and prediction and the mismatch between the prediction model and the true channel model. 
In contrast, most JCEP algorithms outperform non-JCEP ones, which verifies the necessity of joint operation of channel estimation and prediction.

\begin{figure}[!tb]
	\centering
	\subfigure[Channel estimation]{
		\includegraphics[width=0.85\linewidth]{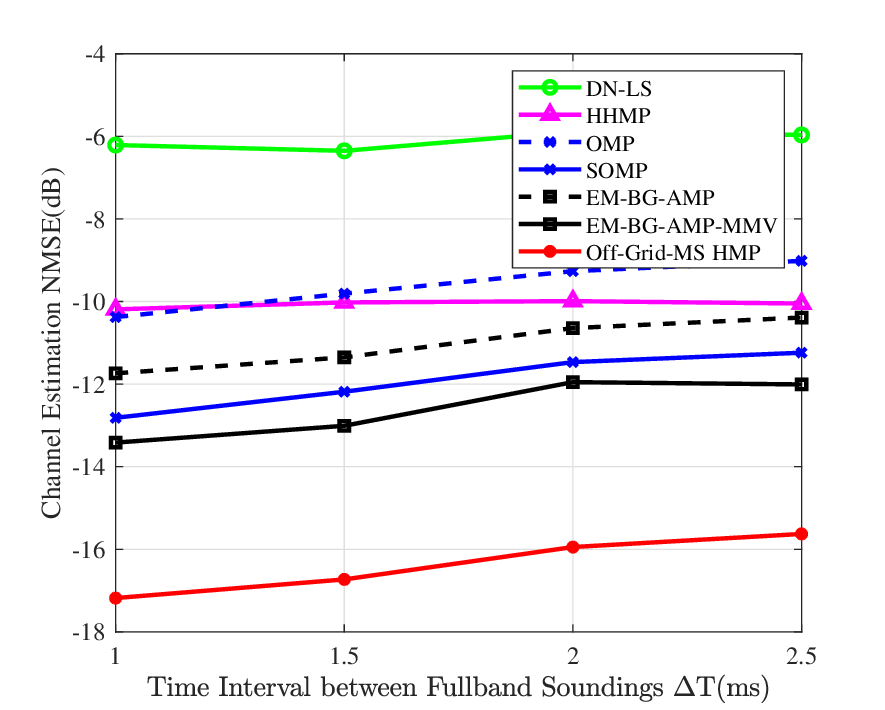}
	}
	\subfigure[Channel prediction]{
		\includegraphics[width=0.85\linewidth]{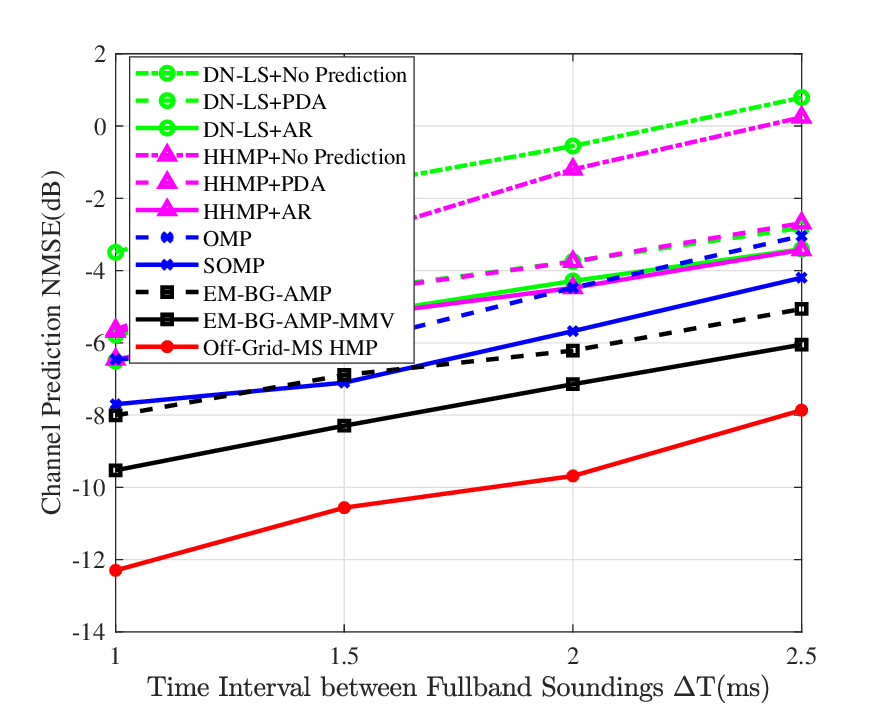}
	}	
	\caption{The performance of JCEP versus $ \Delta T $.}
	\label{fig:simu DeltaT}
\end{figure} 
Since the trend of the performance with the pilot overhead is crucial in the system design, we depict the NMSE of different algorithms versus the time interval between the adjacent fullband soundings $ \Delta T $ at the SNR of 0 dB in Fig.~\ref{fig:simu DeltaT}. 
As shown in Fig. \ref{fig:simu DeltaT}(a), the channel estimation performance of JCEP algorithms declines with increasing time interval $ \Delta T $ because of the utilization of temporal correlation, whereas the time interval $ \Delta T $ has no effect on the estimation accuracy of the non-JCEP algorithms, which independently estimates the channels of different symbols. 
As shown in Fig. \ref{fig:simu DeltaT}(b),  it is evident that the channel prediction performance of all algorithms deteriorates with increased time interval $ \Delta T $ due to the rapid time-varying characteristics of the channel, indicating the pilot overhead as a crucial factor in the CSI acquisition. 
The proposed algorithm consistently outperforms all baselines for any time interval $ \Delta T $, which implies the pilot overhead can be significantly reduced compared to baselines at the same NMSE. 
Specifically, the proposed algorithm can reduce the pilot overhead by over 60\% for -14 dB channel estimation NMSE and by about 50\% for -10 dB channel prediction NMSE compared to the best performance baseline.

\begin{figure}[!tb]
	\centering
	\subfigure[Channel estimation]{
		\includegraphics[width=0.85\linewidth]{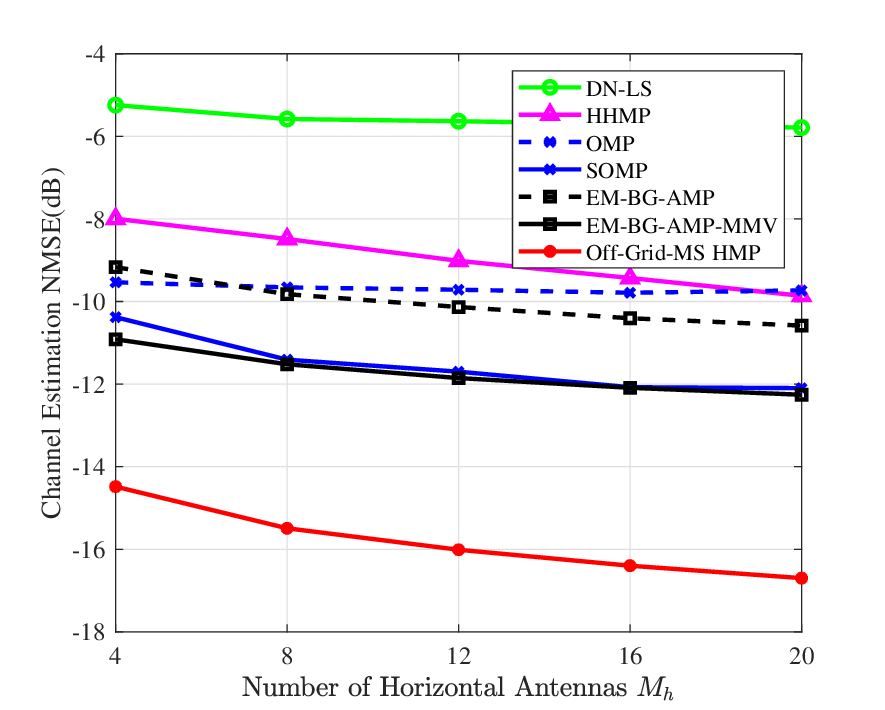}
	}
	\subfigure[Channel prediction]{
		\includegraphics[width=0.85\linewidth]{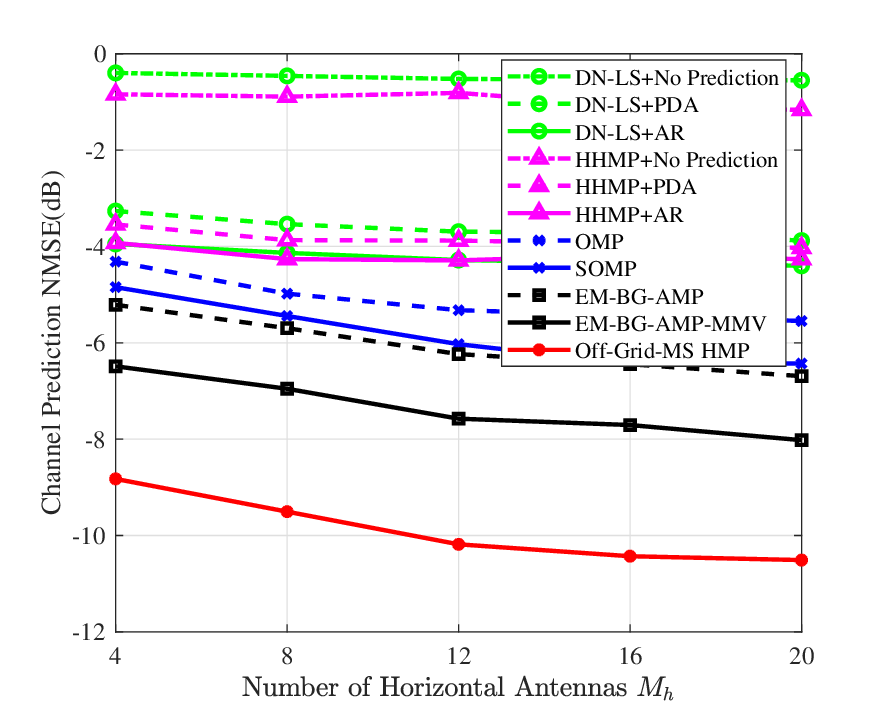}
	}	
	\caption{The performance of JCEP versus $ M_{\text{h}} $.}
	\label{fig:simu M}
\end{figure}
The NMSE versus the number of BS antennas at the SNR of 0 dB is presented in Fig. \ref{fig:simu M}. To directly illustrate the effect of the number of BS antennas, we set $ M_{\text{v}} = 1 $ and adjust the value of $ M_{\text{h}} $ to observe the performance changes of different algorithms. It can be found that the proposed algorithm reaches the best JCEP performance compared with other Baselines at the entire region of $ M_{\text{h}} $, and its performance improves with an increase in the number of BS antennas. Specifically, the performance is proportional to the number of BS antennas in a small-scale antenna array scenario due to its low resolution in the angle domain. In contrast, the performance is minimally influenced by the number of BS antennas in a large-scale case, as its angle-domain resolution is high enough to distinguish the angles from different paths.

\section{Conclusion}
\label{sec6}
In this paper, we investigated the JCEP problem for massive MIMO with FHS.
Firstly, we developed an accurate DAD-domain channel model to mitigate the DFT energy leakage issue caused by low-resolution DAD-domain channels based on DFT in practical scenarios. 
Subsequently, we formulated the JCEP problem with FHS as a generalized MMV problem based on the foundation that the channels of different subbands are i.i.d. 
To achieve efficient JCEP with FHS, we approximated the MMV problem as a BFE minimization problem with hybrid constraints and proposed an efficient Off-Grid-MS HMP algorithm capable of adaptively learning the hyper-parameter. 
In addition, we leveraged the approximations of off-grid matrices to reduce the complexity of hyper-parameter learning. 
Numerical results illustrated that the proposed algorithm could significantly enhance channel estimation and prediction performance compared to state-of-the-art counterparts.

% or


\appendices
\section{Derivation of \eqref{eq:EM off-grid2}}\label{Appendix A}
The optimization object can be written as 
\begin{align}
	\label{eq:Appendix EM alpha}
	&\textstyle\text{E}\Big[
	\sum_{l}\big(
	\prod_{n,m,k}b^{g\mathbf{h}}_{nmkl}\big)\ln \big(\prod_{n,m,k}p(g_{nmkl}\vert\mathbf{h}_{l};\boldsymbol{\omega} )\big)
	\Big] \nonumber\\
	&\textstyle\propto\sum_{l}
	\text{E}\Big[
	-\frac{1}{\epsilon}
	\big\Vert
	\mathbf{g}_{l}-\mathbf{W}(\boldsymbol{\omega}) \mathbf{h}_{l}
	\big\Vert_{2}^{2}
	\Big\vert \prod_{n,m,k}b^{g\mathbf{h}}_{nmkl}
	\Big] \\
	&\textstyle\propto
	\frac{1}{L}\sum_{l}\big\Vert
	\boldsymbol{\mu}^{g}_{l}-\mathbf{W}(\boldsymbol{\omega}) \boldsymbol{\mu}^{h}_{l}
	\big\Vert_{2}^{2} +  \text{Tr}\big\lbrace
	\mathbf{W}(\boldsymbol{\omega})
	\boldsymbol{\Sigma}^{h}
	( \mathbf{W}(\boldsymbol{\omega}) )^{H}
	\big\rbrace. \nonumber
\end{align}

Firstly, we simplify the norm in the first term of \eqref{eq:Appendix EM alpha} as
\begin{align}
	\label{eq:Appendix EM alpha first term}
	&\textstyle
	\big\Vert
	\boldsymbol{\mu}^{g}_{l}-\mathbf{W}(\boldsymbol{\omega}) \boldsymbol{\mu}^{h}_{l}
	\big\Vert_{2}^{2}
	\nonumber\\
	&\textstyle\quad\overset{\text{(a)}}{=}\big\Vert
	\boldsymbol{\mu}^{g}_{l}
	-\mathbf{W}_{\backslash x} \boldsymbol{\mu}^{h}_{l}
	-
	\dot{\mathbf{W}}_{x}
	\text{diag}\lbrace \boldsymbol{\mu}^{h}_{l} \rbrace
	\mathbf{R}_{x}\mathbf{x}
	\big\Vert_{2}^{2} \nonumber\\
	&\textstyle\quad\overset{\text{(b)}}{\propto}
	\mathbf{x}^{T}\mathbf{R}_{x}^{T}
	((
	\dot{\mathbf{W}}_{x}^{H}
	\dot{\mathbf{W}}_{x}
	)^{*}
	\odot
	( \boldsymbol{\mu}^{h}_{l}(\boldsymbol{\mu}^{h}_{l})^{H} ))
	\mathbf{R}_{x}\mathbf{x} 
	\nonumber\\
	&\textstyle\quad\quad -
	2\text{Re}\lbrace
	\text{diag}\lbrace (\boldsymbol{\mu}^{h}_{l})^{*} \rbrace
	\dot{\mathbf{W}}_{x}^{H} 
	( \boldsymbol{\mu}^{g}_{l}
	-
	\mathbf{W}_{\backslash x} \boldsymbol{\mu}^{h}_{l} ) 
	\rbrace^{T} \mathbf{R}_{x}\mathbf{x},
\end{align}
where $ \text{(a)} $ is obtained according to equation $ \text{diag}\lbrace\mathbf{x}\rbrace \mathbf{y}\triangleq\text{diag}\lbrace\mathbf{y}\rbrace \mathbf{x} $, 
and $ \text{(b)} $ is from equation $ \text{diag}^{H}\lbrace \mathbf{x} \rbrace\mathbf{Y}^{H}\mathbf{Y}\text{diag}\lbrace \mathbf{x} \rbrace\triangleq (\mathbf{Y}^{H}\mathbf{Y})^{*}\odot(\mathbf{x}\mathbf{x}^{H}) $. 

Then, the second term in \eqref{eq:Appendix EM alpha} can be simplified as
\begin{align}
	\label{eq:Appendix EM alpha second term}
	&\textstyle\text{Tr}\big\lbrace
	\mathbf{W}(\boldsymbol{\omega})
	\boldsymbol{\Sigma}^{h}
	( \mathbf{W}(\boldsymbol{\omega}) )^{H}
	\big\rbrace \nonumber\\
	&\textstyle\quad\overset{\text{(a)}}{\propto}\text{Tr}\big\lbrace
	( \dot{\mathbf{W}}_{x}\text{diag}\lbrace\mathbf{R}_{x}\mathbf{x}\rbrace )^{H}
	( \dot{\mathbf{W}}_{x}\text{diag}\lbrace\mathbf{R}_{x}\mathbf{x}\rbrace )
	\boldsymbol{\Sigma}^{h}
	\big\rbrace \nonumber\\
	&\textstyle\qquad + 2\text{Re}\big\lbrace
	\text{Tr}\big\lbrace
	( \dot{\mathbf{W}}_{x}\text{diag}\lbrace\mathbf{R}_{x}\mathbf{x}\rbrace )^{H}
	\mathbf{W}_{\backslash x}
	\boldsymbol{\Sigma}^{h}
	\big\rbrace
	\big\rbrace \nonumber\\
	&\textstyle\quad\overset{\text{(c)}}{\propto}
	\mathbf{x}^{T}\mathbf{R}_{x}^{T}
	((
	\dot{\mathbf{W}}_{x}^{H}
	\dot{\mathbf{W}}_{x}
	)^{*}
	\odot
	\boldsymbol{\Sigma}^{h})
	\mathbf{R}_{x}\mathbf{x} \nonumber\\
	&\textstyle\qquad  + 
	2\text{Re}\big\lbrace
	\text{diag}\big\lbrace
	\dot{\mathbf{W}}_{x}^{H}
	\mathbf{W}_{\backslash x}
	\boldsymbol{\Sigma}^{h}
	\big\rbrace
	\big\rbrace^{T} \mathbf{R}_{x}\mathbf{x} , 
\end{align}
where $ \text{(a)} $ is from equation $ \text{Tr}\lbrace\mathbf{X}\mathbf{Y}\rbrace\triangleq\text{Tr}\lbrace\mathbf{Y}\mathbf{X}\rbrace $, 
and $ \text{(b)} $ is obtained based on equation $ \text{Tr}\lbrace \text{diag}^{H}\lbrace \mathbf{x} \rbrace \mathbf{X}\text{diag}\lbrace \mathbf{y} \rbrace \mathbf{Y}^{T} \rbrace \triangleq \mathbf{x}^{H}(\mathbf{X}\odot \mathbf{Y}) \mathbf{y} $ and the definition of trace. 

Finally, since the matrix $ \mathbf{Y} $ is semi-definite, and the vector $ \mathbf{x} $ is real, we can derive \eqref{eq:EM off-grid2} based on \eqref{eq:Appendix EM alpha first term} and \eqref{eq:Appendix EM alpha second term}.

%\section{XXX}

%\section*{Acknowledgment}

\ifCLASSOPTIONcaptionsoff
  \newpage
\fi



\bibliographystyle{IEEEtran}
\bibliography{IEEEabrv,reference}

\end{document}